\documentclass[aps,pra,showpacs]{revtex4}

\usepackage[utf8x]{inputenc}
\usepackage{amsmath}
\usepackage{amsthm}
\usepackage{amssymb}
\usepackage{amsfonts}
\usepackage{graphics}
\usepackage[pdftex]{graphicx}
\usepackage[normal]{subfigure}
\usepackage{color}

\begin{document}

\title{Discrete-time Quantum Walks in random artificial Gauge Fields}
\author{G. Di Molfetta$^*$}
\author{F. Debbasch$^*$}
\affiliation{$^*$ LERMA, Observatoire de Paris, PSL Research University, CNRS, Sorbonne Universités, UPMC Univ. Paris 6, UMR 8112, F-75014, Paris France}

\date{\today}

\begin{abstract}
Discrete-time quantum walks (DTQWs) in random artificial electric and gravitational fields are studied analytically and numerically. The analytical computations are carried by a new method which allows a direct exact analytical determination of the equations of motion obeyed by the average density operator. It is proven that randomness induces decoherence and that the quantum walks behave asymptotically like classical random walks. Asymptotic diffusion coefficients are computed exactly. The continuous limit is also obtained and discussed.
\end{abstract}


\maketitle

\section{Introduction}

Discrete time quantum walks (DTQWs) are simple formal analogues of classical random walks. They were first considered by Feynmann in \cite{FeynHibbs65a}, and then introduced in greater generality in \cite{ADZ93a} and \cite{Meyer96a}. They have been realized experimentally  \cite{Schmitz09a, Zahring10a, Schreiber10a, Karski09a, Sansoni11a, Sanders03a, Perets08a} and are important in many fields, ranging from fundamental quantum physics \cite{Perets08a, var96a} to quantum algorithmics \cite{Amb07a, MNRS07a}, solid state physics \cite{Aslangul05a, Bose03a, Burg06a, Bose07a} and biophysics \cite{Collini10a, Engel07a}.

It has been shown \cite{DMD12a, DMD13b,DMD14} recently that several DTQWs on the line admit a continuous limit identical to the propagation of a Dirac fermion in artificial electric and gravitational fields. These DTQWs are thus simple discrete models of quantum propagation in artificial gauge fields. Here, we consider artificial gauge fields which depend randomly on time and investigate analytically and numerically how this randomness influences quantum propagation. 
%To the best of our knowledge, all previous work on random DTQWs was done by performing %averages numerically. 
The analysis presented in this article is based on a 
%new method which allows a 
direct analytical computation of the exact evolution equation obeyed by the average density operator. This presents several advantages. First, the average dynamics is thus known exactly, without the noise inherent in any numerical evaluation of averages. Second, knowing the exact average equations of motion makes it possible to study the average dynamics analytically.
%, and this would be naturally impossible if the average dynamics were only known numerically.  
Finally, simulating directly the exact analytical equations of the average dynamics offers a significant gain in computation time over alternative methods where the average evolution is determined by simulating successively a large number of realizations of the random  DTQWs.

Random DTQWs have already been studied by several authors (see for example \cite{Ken07a, vieira14, brun2003, Werner11, Cedzich12,Joye11}) , but the influence of random gauge fields has never been the object of specific analytical computations. In particular, exact expressions of the asymptotic density profiles as functions of the randomness caracteristics have never been computed. 
%Our results confirm previous findings \cite{Ken07a,vieira14} which suggest that randomness in time causes the walk to loose coherence. Indeed, 
Our main results are (i) DTQWs interacting with artificial gauge fields which are random in time decohere and behave asymptotically like classical random walks (ii) 
%We also compute the
the asymptotic density profiles of the DTQWs 
%and prove they 
are Gaussian and we give exact analytical expressions of the asymptotic diffusion coefficients as functions of the noise amplitude which generates the randomness. We also support 
%(iii) that the 
%%We obtain exact analytical expressions for the 
%asymptotic diffusion coefficients can be computed exactly analytically and support 
all results by direct numerical simulations of the average dynamics and finally discuss the continuous limits of the DTQWs interacting with random artificial gauge fields.

\section{A family of DTQWs coupled to artificial electric and gravitational fields}

\subsection{Wave-function evolution}

\subsubsection{In physical space}

We consider discrete time quantum walks in one space dimension driven by a 
time-dependent quantum coin acting on a two-dimensional Hilbert space $\mathcal H$.
The walks are defined by the following finite difference equations, valid for all $(j, m) 
\in \mathbb{N}  \times \mathbb{Z}$:  
\begin{equation}
\begin{bmatrix} \psi^{L}_{j+1, m }\\ \psi^{R}_{j+1, m } \end{bmatrix} \  = 
{\mathcal B}\left(\theta_{j} ,\xi_{j} \right)
 \begin{bmatrix} \psi^{L}_{j, m+1} \\ \psi^{R}_{j, m-1} \end{bmatrix},
\label{eq:defwalkdiscr}
\end{equation}
where 
\begin{equation}
 {\mathcal B}(\theta ,\xi) = 
\begin{bmatrix}  e^{i\xi} \cos\theta &  i \sin\theta\\ i \sin\theta &  e^{-i\xi} \cos\theta
 \end{bmatrix}.
\label{eq:defB}
\end{equation} 

The operator represented by the matrix $\mathcal B$ is in $SU(2)$ and $\theta$ and $\xi$ are two of the three Euler angles.
The index $j$ labels instants and takes all positive integer values. The index $m$ labels spatial points. We choose to work on the circle and impose periodic boundary conditions. We thus introduce a strictly positive integer $M$ and restrict $m$ to all integer values between $-M$ and $+M$ {\sl i.e.} $m \in \mathbb Z_M$. Results pertaining to DTQWs on the infinite line can be recovered by letting $M$ tend to infinity. 

For each instant $j$ and each spatial point $m$, the wave function $\Psi_{jm} = \psi^L_{jm} b_L + \psi^R_{jm} b_R = \psi^a_{jm} b_a, a \in\left\{L, R \right\}$, has two components $\psi^L_{jm} $ and $\psi^R_{jm}$ on the spin basis $(b_L, b_R)$ and these code for the probability amplitudes 
of the particle jumping towards the left or towards the right. Note that the spin basis is interpreted as being independent of $j$ and $m$.
For a given initial condition, the set of angles $\left\{ \theta_{j},\xi_{j},  j \in \mathbb{N} \right\}$ completely defines the walks and is arbitrary.

It has been proven in \cite{DMD12a, DMD13b,DMD14} that walks from this family are models of Dirac fermions coupled to artificial electric and gravitational fields. Details can be found in these references and in the first appendix to the present article.

\subsubsection{In Fourier space}

A practical tool to study quantum walks on the discrete circle is the discrete Fourier transform (DFT). Let $(A_m)_{m \in \mathbb Z_M}$ be an arbitray sequence of complex numbers defined on the discrete circle. The DFT of this sequence is the sequence $({\hat A}_{k_n})_{n \in \mathbb Z_M}$ defined by
\begin{equation}
{\hat A}_{k_n} = \sum_{m = -M}^{+M} A_m \exp \left( i k_n m\right)
\end{equation}
with $k_n = 2 n \pi/(2 M +1)$, $n \in \mathbb Z_M$. The original sequence can be recovered from its DFT by the relation:
\begin{equation}
{A}_{m} = \frac{1}{2M + 1}\sum_{n = -M}^{+M} {\hat A}_{k_n} \exp \left(- i k_n m\right).
\end{equation}
For infinite $M$ {\sl i.e.} DTQWs on the infinite line, the DFT of an infinite sequence $(A_m)_{m \in \mathbb Z}$ becomes a function
\begin{equation}
{\hat A}(k) = \sum_{m \in \mathbb Z} A_m \exp \left( i k m\right)
\end{equation}
defined for $k \in (- \pi, \pi )$ and the inverse relation reads:
\begin{equation}
{A}_{m} = \frac{1}{2\pi} \int_{- \pi}^{\pi} {\hat A}(k) \exp \left(- i k m\right) dk.
\end{equation}
In Fourier space on the infinite line, the evolution equation (\ref{eq:defwalkdiscr}) transcribes into
\begin{equation}
\begin{bmatrix} {\hat \psi}^{L}_{j+1}(k)\\ {\hat \psi}^{R}_{j+1}(k) \end{bmatrix} \  = 
{\mathcal C}(\theta_j, \xi_j, k)
 \begin{bmatrix} {\hat \psi}^{L}_{j}(k) \\ {\hat \psi}^{R}_{j}(k) \end{bmatrix}
\label{eq:defwalkdiscr2}
\end{equation}
where 
\begin{equation}
{\mathcal C}(\theta_j, \xi_j, k) = 
\begin{bmatrix}  e^{i\xi} \cos\theta e^{- i k}&  i \sin\theta e^{+ i k}\\ i \sin\theta e^{- i k}&  e^{-i\xi} \cos\theta e^{+ i k}
 \end{bmatrix}
\end{equation}
for all $k \in (- \pi, \pi)$.

\subsection{Density operator evolution}

\subsubsection{In physical space}

The walks can also be described using the density operator $\rho = \Psi^* \otimes \Psi$. We introduce the basis $v_1 = b_L \otimes b_L$, 
$v_2 = b_L \otimes b_R$, $v_3 = b_R \otimes b_L$, $v_4 = b_R \otimes b_R$ and represent $\rho$ by its components %$\rho^a_{j, m, m'}$, $a \in \{1, 2, 3, 4\}$, 
on this basis
{\sl i.e.} by the quantities
$\rho^{a b}_{j,m,m'} = \psi^{b *}_{j m'} \psi^{a}_{j m}$, $\{a, b\} \in \{L, R\}^2$.
%with components  
%%$\rho_{j,m,m'}$ = $\Psi_{j,m}^*\Psi_{j,m'}$ and its matrix representation in the basis %$\{b_+,b_-\}$,  
%$\rho^{a b}_{j,m,m'} = \psi^{b *}_{j m'} \psi^{a}_{j m}$, $\{a, b\} \in \{L, R\}^2$.
%%$ b_\nu^* \rho_{j,m,m'} b_\mu$.
%%\\
Equation (\ref{eq:defwalkdiscr}) delivers:
\vspace{0.2cm}
\begin{equation}
\begin{bmatrix}
\rho^{LL}_{j+1,m,m'}\\ \rho^{LR}_{j+1,m,m'} \\ \rho^{RL}_{j+1,m,m'}\\ \rho^{RR}_{j+1,m,m'}
\end{bmatrix} =\mathcal{Q}\left( \theta_{j},\xi_{j}\right) \begin{bmatrix}
\rho^{LL}_{j,m+1,m'+1}\\ \rho^{LR}_{j,m+1,m'-1} \\ \rho^{RL}_{j,m-1,m'+1} \\ \rho^{RR}_{j,m-1,m'-1}
\end{bmatrix} 
\label{eq:defdens}
\end{equation}
where  
\begin{equation}
\mathcal{Q} (\theta, \xi)=
\begin{bmatrix}
c^2 & -ics\,  e^{+i \xi} & +ics\,  e^{- i \xi} & s^2 \\
-ics\,  e^{+i \xi} & c^2 e^{+2i \xi} & s^2 & +ics\,  e^{+i \xi} \\
+ics\,  e^{- i \xi} & s^2 & c^2 e^{-2i \xi} & - ics\,  e^{- i \xi} \\
s^2 &  +ics\,  e^{+ i \xi} & -ics\,  e^{- i \xi} & c^2 
\end{bmatrix},
\end{equation}
with $c= \cos\theta$ and $s = \sin\theta$. The probability to find the walk at time $j$  at point $m$ is $N_{jm} = \rho^{LL}_{j,m,m}+ \rho^{RR}_{j,m,m}$ and the sum $\sum_m N_{jm}$ is independent of $j$ {\sl i.e.} it is conserved by the walk. Contrary to equation (\ref{eq:defwalkdiscr}), equation (\ref{eq:defdens}) can be used to describe walks with initial conditions which are not pure states. Equation (\ref{eq:defdens}) is thus more general than (\ref{eq:defwalkdiscr}). 

%Since the operator $\mathcal B$ is unitary, the operator $\mathcal Q$ governing the evolution of $\rho$ is also unitary, as can be checked by a straightforward computation.

%It has been proven in \cite{DMD13b,DMD14} that the walks defined by (\ref{eq:defwalkdiscr}) can be interpreted as the transport of a Dirac fermion in artificial electric and gravitational fields generated by the time-dependance of the angles $\theta$ and $\xi$. 

\subsubsection{In Fourier space}
%
%A practical tool to study quantum walks on the discrete circle is the discrete Fourier transform (DFT). Let $(A_m)_{m \in \mathbb Z_M}$ be an arbitray sequence of complex numbers defined on the discrete circle. The DFT of this sequence is the sequence $({\hat A}_{k_n})_{n \in \mathbb Z_M}$ defined by
%\begin{equation}
%{\hat A}_{k_n} = \sum_{m = -M}^{+M} A_m \exp \left( i k_n m\right)
%\end{equation}
%with $k_n = 2 n \pi/(2 M +1)$, $n \in \mathbb Z_M$. The original sequence can be recovered from its DFT by the relation:
%\begin{equation}
%{A}_{m} = \frac{1}{2M + 1}\sum_{n = -M}^{+M} {\hat A}_{k_n} \exp \left(- i k_n m\right).
%\end{equation}
%For infinite $M$ {\sl i.e.} DTQWs on the infinite line, the DFT of an infinite sequence $(A_m)_{m \in \mathbb Z}$ becomes a function
%\begin{equation}
%{\hat A}(k) = \sum_{m \in \mathbb Z} A_m \exp \left( i k m\right)
%\end{equation}
%defined for $k \in (- \pi, \pi )$ and the inverse relation reads:
%\begin{equation}
%{A}_{m} = \frac{1}{2\pi} \int_{- \pi}^{\pi} {\hat A}(k) \exp \left(- i k m\right) dk.
%\end{equation}

%At any instant $j$, the density operator $\rho$ depends on two space variables $m$ and $m'$. 
Consider now, for any instant $j$, the double DFT of the density operator $\rho_{j, m, m'}$, which we denote by 
%${\hat \rho}_{j, k, k'}$
${\hat \rho}_{j} (k, k')$ 
or, alternately,  
%${\hat \rho}_{j, K, p}$
${\hat \rho}_{j} (K, p)$
 where $K = (k + k')/2$ is conjugate to $m + m'$ and $p = (k' - k)/2$ is conjugate to $m' -m$. For DTQWs on the infinite line, the range of both $K$ and $p$ is $( - \pi, + \pi)$. The DFT of the density operator obeys ${\hat \rho}_{j + 1} (K, p) = {\mathcal R}\left( \theta_{j},\xi_{j}, K, p\right) {\hat \rho}_{j} (K, p)$ with
\vspace{0.2cm}
%\begin{equation}
%\begin{bmatrix}
%{\hat \rho}^{LL}_{j+1, K, p}\\ {\hat \rho}^{LR}_{j+1, K, p} \\ {\hat \rho}^{RL}_{j+1, K, p}\\ {\hat \rho}^{RR}_{j+1, K, p}
%\end{bmatrix} =\mathcal{R}\left( \theta_{j},\xi_{j}, K, p\right) \begin{bmatrix}
%{\hat \rho}^{LL}_{j, K, p}\\ {\hat \rho}^{LR}_{j, K, p} \\ {\hat \rho}^{RL}_{j, K, p} \\ {\hat \rho}^{RR}_{j, K, p}
%\end{bmatrix} 
%\label{eq:defdens}
%\end{equation}
%where  
\begin{equation}
\mathcal{R}(\theta, \xi, K, p)=
\begin{bmatrix}
c^2\,  e^{2iK} & -ics\,  e^{+i \xi}\,  e^{-2ip} & +ics\,  e^{- i \xi}\, e^{+2ip} & s^2\, e^{-2iK}\\
-ics\,  e^{+i \xi} \,  e^{2iK} & c^2 e^{+2i \xi} \,  e^{-2ip} & s^2 \, e^{+2ip} & +ics\,  e^{+i \xi} \, e^{-2iK}\\
+ics\,  e^{- i \xi} \,  e^{2iK} & s^2 \,  e^{-2ip} & c^2 e^{-2i \xi}  \, e^{+2ip} & - ics\,  e^{- i \xi} \, e^{-2iK}\\
s^2 \,  e^{2iK} &  +ics\,  e^{+ i \xi} \,  e^{-2ip} & -ics\,  e^{- i \xi} \, e^{+2ip} & c^2 \, e^{-2iK}
\end{bmatrix}.
\label{eq:RthetaxiKP}
\end{equation}
%Note that $\mathcal R$, like $\mathcal Q$, is unitary.

Note that the operator $\mathcal R$ governing the evolution of ${\bar \rho}$ is unitary. This can be checked by a straightforward computation and it is a direct consequence of the unitarity of the operator $\mathcal B$.

%The indices $c$ and $d$ here refer to the basis $\{v_1, v_2, v_3, v_4\}$ and take any integer value between $1$ and $4$.

%In this new basis, the matrix representing the operator $\mathcal{R}(\theta, \xi, K, p)$ reads:

\section{Randomizing the fields and averaging the dynamics}

\subsection{Randomizing the fields}

The Hadamard walk corresponds to $\xi = \xi_H= \pi/2$ and $\theta = \theta_H = \pi/4$; since these angles are constant, the Hadamard walk describes propagation in the absence of electric and gravitational field \cite{DMD13b,DMD14}. We now consider situations where one of the angles $\xi$ and $\theta$ does depend on time and fluctuates around its Hadamard value.
More precisely, we consider two cases.
%We choose to consider two typical cases, which can be viewed as randomly perturbed (or noisy) Hadamard walks. 
Case 1 corresponds to $\theta = \theta_H = \pi/4$ and $\xi$ chosen randomly at each time-step with uniform probability law in the interval $(\pi/2 - \sigma/2, \pi/2 + \sigma/2)$, where $\sigma \in (0, 2 \pi)$ is a fixed {\sl i.e.} $j$-independent positive real number. As proven in \cite{DMD12a,DMD13b,DMD14} and detailed in the first appendix to the present article, a time-dependent $\theta$ is equivalent to a space-time metric whose purely spatial part depends on time, and such a metric represents a time-dependent relativistic gravitational field. Case 2 corresponds to $\xi = \xi_H = \pi/2$ and $\theta$ chosen randomly at each time-step with uniform probability law in the interval $(\pi/4 - \sigma/2, \pi/4 + \sigma/2)$. As proven in \cite{DMD14}, a time-dependent $\xi$ is equivalent to a time-dependent `vector' potential, which represents a time-dependent electric feld.  
%We restrict $\sigma$ to $(0, 2 \pi)$, so that $\xi$ and $\theta$ describe intervals of lengths $\sigma \le 2 \pi$. 

Thus, in each case, a realization of the random gauge field is determined by a sequence $\omega = \left(\omega_1, \omega_2, ...\right)$ of independent random variables, where $\omega_j$ represents the value of the random angle $\theta$ or $\xi$ at time $j$. If one follows the walk till time $N$, the relevant random sequence is the $N$-uple $\omega^N = \left(\omega_1, \omega_2, ..., \omega_N\right)$. For each value of $\sigma$ and each instant $j$, $\omega_j$ is uniformly distributed in the interval  $I_\sigma = (\omega_H - \sigma/2, \omega_H + \sigma/2)$ centered on the Hadamard value $\omega_H$. The probability density of $\omega_j$ in this interval is thus simply $p_\sigma (\omega_j) = 1/\sigma$ and is independent of both $\omega_j$ and $j$. The probability density for $\omega^N = \left(\omega_1, \omega_2, ..., \omega_N\right)$ in $I_\sigma^N$ is therefore $P_\sigma (\omega^N) = \Pi_{j = 1}^N p_\sigma (\omega_j)  = 1/\sigma^N$ and is independent of $\omega^N$.

\subsection{Averaging the dynamics}

At fixed initial condition $\rho_0$ and for each time $N$,
%and for each realization $\omega^N$ of the random gauge field, 
the density operator $\rho_N$ at time $N$ depends on the realization $\omega^N$ of the random angle up to time $N$. 
At fixed initial condition, the easiest way to compute statistical averages over $\omega^N$ is to first compute the statistical average $\bar{\rho}_N$ of the density operator over $\omega^N$:
\begin{eqnarray}
\bar{\rho}_N & = & \int_{I_\sigma^N} \rho_N(\omega^N) P_\sigma (\omega^N) d\omega^N \nonumber \\
& = & \int_{I_\sigma^N} \rho_N(\omega_1, ..., \omega_N) p_\sigma(\omega_1) ... p_\sigma(\omega_N) d\omega_1...d\omega_N \nonumber \\
& = & \int_{I_\sigma^N} \rho_N(\omega_1, ..., \omega_N) \frac{1}{\sigma^N}\, d\omega_1...d\omega_N.
\end{eqnarray}
Let us work in Fourier space. One can then write, for any realization $\omega^N = (\omega_1, ..., \omega_N)$ of the random angle up to time $N$:
%the DFT ${\hat \rho}_j$ of density operator $\rho_j$ at time $j$ is given by:
\begin{eqnarray}
{\hat \rho}_N (\omega_1, ..., \omega_N) & = & {\mathcal R}(\omega_N) {\hat \rho}_{N-1}(\omega_{1}, ..., \omega_{N-1}) \nonumber \\
& = & {\mathcal R}(\omega_N) ... {\mathcal R}(\omega_1) {\hat \rho}_{0},
\end{eqnarray} 
where the variables $K$ and $p$ 
%and the indices $L$ and $R$ 
have been omitted for clarity reasons.
Since the $\omega_j$'s are statistically independent of each other and are identically distributed, one obtains:
\begin{equation}
{\hat {\bar \rho}}_N = {\bar {\mathcal R}}^N  {\hat {\bar \rho}}_{0}
\end{equation}
%where
%\begin{eqnarray}
%{\hat {\bar \rho}}_N & = & {\bar {\mathcal R}} {\bar {\hat \rho}}_{j-1} \nonumber \\
%& = & {\bar {\mathcal R}}^j  {\bar {\hat \rho}}_{0}
%\end{eqnarray} 
where ${\bar {\mathcal R}}$ is the statistical average of the evolution operator $\mathcal R$ over the random angle $\omega = \theta$ or $\xi$ (the other angle being fixed to its Hadamard value):
\begin{eqnarray}
{\bar {\mathcal R}} (K, p, \sigma) & = & \int_{I_\sigma} {\mathcal R} (\omega, K, p) p_\sigma (\omega) d\omega \nonumber \\
{\bar {\mathcal R}} (K, p, \sigma) & = & \int_{I_\sigma} {\mathcal R} (\omega, K, p) \frac{1}{\sigma}\,  d\omega.
\end{eqnarray}

The average evolution ${\bar {\mathcal R}}$ is thus a function of $(K, p)$ and of the noise parameter $\sigma$ and
%This average operator 
can be computed analytically from (\ref{eq:RthetaxiKP}). It determines the evolution of the average density operator completely and, therefore, the average transport. Since everything that follows pertains only to the average transport, we simplify the notation by droping the bar on the letter $\rho$ and the density operator of the averaged transport will now be designated simply by $\rho$.
 
%The averaged density operator can thus be computed at any time once the averaged evolution operator ${\bar {\mathcal R}}$ is known. 

A direct computation from (\ref{eq:RthetaxiKP}) leads to the following exact expressions for the components of $\bar{\mathcal  R}$ in the basis $\{v_1, v_2, v_3, v_4\}$ for case 1 (random electric field) and case 2 (random gravitational field):
\begin{equation}
{\bar {\mathcal R}}^e(K, p, \sigma)= \frac{1}{2} 
\begin{bmatrix}
e^{2iK} & \text{sinc}(\sigma/2)\, e^{-2ip} & \text{sinc}(\sigma/2) \, e^{+2ip} &  e^{-2iK}\\
\text{sinc}(\sigma/2) \,  e^{2iK} & -  \text{sinc}(\sigma) \, e^{-2ip} & e^{+2ip} & - \text{sinc}(\sigma/2) \, e^{-2iK}\\
\text{sinc}(\sigma/2) \,  e^{2iK} &  e^{-2ip} & - \text{sinc}(\sigma) \, e^{+2ip} & - \text{sinc}(\sigma/2) \, e^{-2iK}\\
e^{2iK} & - \text{sinc}(\sigma/2) \,  e^{-2ip} & -\text{sinc}(\sigma/2) \, e^{+2ip} & e^{-2iK}
\end{bmatrix},
\end{equation}
and
 \begin{equation}
{\bar {\mathcal R}}^g(K, p, \sigma) = \frac{1}{2} 
\begin{bmatrix}
 e^{2iK} &  \text{sinc}(\sigma) \, e^{-2ip} & \text{sinc}(\sigma) \, e^{+2ip} &  e^{-2iK} \\
 \text{sinc}(\sigma) \,  e^{2iK} & -  e^{-2ip} & e^{+2ip} & - \text{sinc}(\sigma) \, e^{-2iK}\\
\text{sinc}(\sigma) \,  e^{2iK} &  e^{-2ip} & - e^{+2ip} & - \text{sinc}(\sigma) \, e^{-2iK}\\
e^{2iK} & - \text{sinc}(\sigma) \,  e^{-2ip} & -\text{sinc}(\sigma) \, e^{+2ip} & e^{-2iK}
\end{bmatrix},
\end{equation}
%
%\frac{1}{2} \left(
%\begin{array}{cc}
%\begin{array}{cc}
% 1 &  \text{sinc}(\sigma) \\
%  \text{sinc}(\sigma) & -1  \\
%\end{array}
% & 
%\begin{array}{cc}
% \text{sinc}(\sigma) & 1 \\
% 1  & -\text{sinc}(\sigma) \\
%\end{array}
% \\
%
%\begin{array}{cc}
% \text{sinc}(\sigma) & 1  \\
%  1   & -\text{sinc}(\sigma)\\
%\end{array}
%& 
%\begin{array}{cc}
% -1  & - \text{sinc}(\sigma)\\
% - \text{sinc}(\sigma)& 1 \\
%\end{array}
%\\
%\end{array}
%\right)\end{equation}

It proves convenient for all subsequent computations to change basis in $\rho$ space and introduce the new vectors $u_1 = v_1 + v_4$, $u_2 = v_1 - v_4$, $u_3 = v_2 + v_3$, $u_4 = v_2 - v_3$.
%$u_1 = b_L \otimes b_L + b_R \otimes b_R$, $u_2 =  b_L \otimes b_L - b_R \otimes b_R$, $u_3 = b_L \otimes b_R + b_R \otimes b_L$, $u_1 = b_L \otimes b_R - b_R \otimes b_L$. 
In this new basis, the components of 
%Taking $\{u_1, u_2, u_3, u_4 \}$ as default basis in $\rho$ space, we simplify all notations and stop distinguishing explicitly between an intrinsic object, like a vector or an operator, and its components in this basis. With this convention, the operators 
${\bar {\mathcal R}}^e(K, p, \sigma)$ and ${\bar {\mathcal R}}^g(K, p, \sigma)$ read:

\begin{equation}
\displaystyle{
{\bar {\mathcal R}}^e(K, p, \sigma) = 
\begin{bmatrix}
\cos(2K) & i \sin (2K) & 0  & 0 \\
0 & 0  & \text{sinc}(\sigma/2) \cos (2p)& -  i\,  \text{sinc}(\sigma/2) \sin (2p) \\
i \, \text{sinc}(\sigma/2) \sin(2K) &  \text{sinc}(\sigma/2) \cos(2K)  & \frac{ (1 -  \text{sinc}(\sigma))}{2} \cos (2p)& i \frac{( \text{sinc}(\sigma) - 1)}{2} \sin(2p)\\
0 & 0 & i  \frac{(1 +  \text{sinc}(\sigma)) }{2}\sin(2p) & - \frac{( \text{sinc}(\sigma) + 1)}{2}  \cos(2p)
\end{bmatrix},}
\end{equation}

%\begin{equation}
%{\bar {\mathcal R}}^g(K, p, \sigma) = 
%\begin{bmatrix}
%\cos(2K) & 0 & 2 i \text{sinc}(\sigma) \sin (K)  &  \\
%2 i \sin(K) &  0 & \text{sinc}(\sigma) \cos (2K)& 0\\
%0 &  \text{sinc}(\sigma) \cos (p) & 0 & i \sin(p)\\
%0 & - i \text{sinc}(\sigma) \sin(p) & 0 & - \cos(p)
%\end{bmatrix},
%\end{equation}

\begin{equation}
{\bar {\mathcal R}}^g(K, p, \sigma) = 
\begin{bmatrix}
\cos(2K) & i \sin (2K) & 0  & 0 \\
0 & 0  & \text{sinc}(\sigma) \cos (2p)& -  i\,  \text{sinc}(\sigma) \sin (2p) \\
i \, \text{sinc}(\sigma) \sin(2K) &  \text{sinc}(\sigma) \cos(2K)  &  0 & 0\\
0 & 0 & i \sin(2p) & -  \cos(2p)
\end{bmatrix},
\end{equation}

We choose as initial condition the pure state defined by $\Psi_{j = 0, m = 0} = (b_L + i b_R)/\sqrt{2}$
%\begin{equation}
%\Psi_{j = 0, m = 0} = 
%\begin{bmatrix} 1\\ i \end{bmatrix} 
%\end{equation}
and $\Psi_{j = 0, m} = 0$ if $m \ne 0$.
This state corresponds to the density operator $\rho_{j = 0, m= 0, m' = 0} = \left( b_L\otimes b_L + b_R \otimes b_R + i (b_R \otimes b_L - b_L \otimes b_R)\right)/2 = (u_1 - i u_4)/2$
%\begin{equation}
%\rho_{j = 0, m =0, m' = 0} = \begin{bmatrix} 1 \\i \\-i \\1
%\end{bmatrix}
%\end{equation}
%%\b2egin{pmatrix}
%% 1 & i \\ -i & 1
%% \end{pmatrix}
%%\end{equation}
and
$\rho_{j = 0, m, m'} = 0$ if $m \ne 0$
or $m' \ne 0$. In Fourier space, ${\hat \rho}_{j = 0} (K, p) =  (u_1 - i u_4)/2$
%\begin{equation}
%{\hat \rho}_{j = 0, K, p} = \begin{bmatrix} 1 \\i \\-i \\1
%\end{bmatrix}
%\end{equation}
for all $K$ and $p$.

For any realization of the noise {\sl i.e.} for any given value of $\omega$, the initial pure state evolves by the DTQW into a pure state. But the average evolutions descibed by ${\bar {\mathcal R}}^e$ and ${\bar {\mathcal R}}^g$ both transform the initial pure state into a superposition. However, the average transport is symmetrical around the origin, as is the classical Hadamard walk generated from the same initial condition.

Let us finally stress that, contrary to the operator ${\mathcal R}$ governing the unaveraged transport, the operators ${\mathcal R}^{e/g}$ governing the averaged transport are {\sl not} unitary. This loss of unitarity generates  qualitative differences between the unaveraged and the averaged transport. In particular, the averaged transport looses quantum coherence and is asymptotically diffusive. These two important consequences of the averaging process are analyzed in the remaining sections of this article.

%This initial condition generates in each case a symetric averaged transport around the origin. For these transports, the density operator at even times vanishes at all uneven positions $m = 2m' +1$, $m' \in \mathbb Z$, and the density operator at uneven times vanishes at all even positions $m = 2 m'$. To study this transport dynamics, it is thus practically easier to work on a reduced problem obtained form the original one by considering only
%half of the time steps and half of the spatial positions. This procedure is by no means necessary but it seems quite natural, at least from a numerical point of view, and it is also in tune with the continuous limit procedure retained in \cite{ } to study walks coupled to artificial gravitational fields. 

\section{Qualitative description of average transport}
\label{sec;Qualit}
\begin{figure}
\center
\includegraphics[width=0.90\columnwidth]{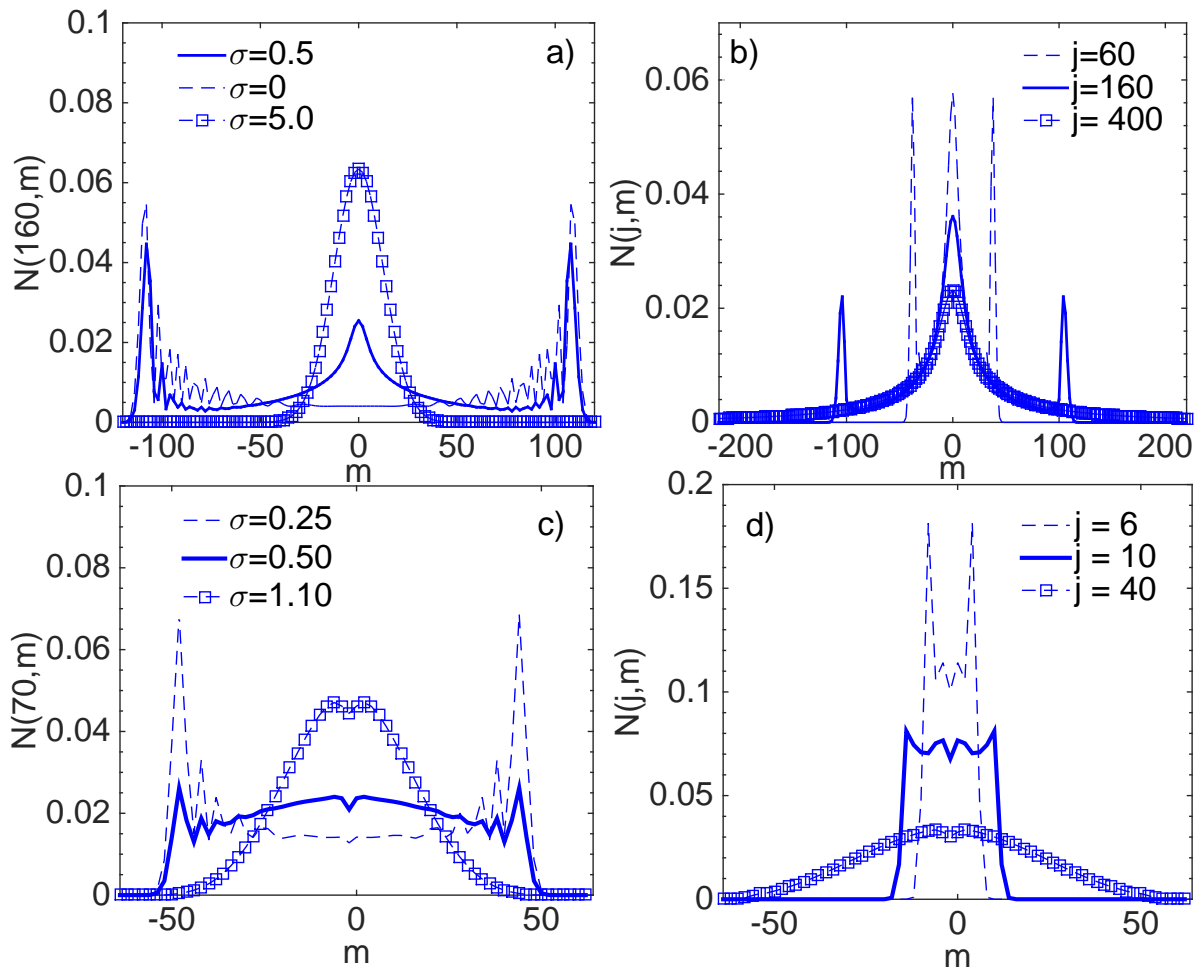}
\caption{(Color online) (left) Probability profile of the average transport in a random $\xi$-field (a) and $\theta$-field (c)  vs grid point m at $j=160$ (a) and  $j=70$ (c) for different values of the noise parameter $\sigma$. 
(right) Probability profile of the average transport in a random $\xi$-field (b) and $\theta$-field (d)  vs grid point m for $\sigma= 0.5$ (d) and $\sigma=0.8$ (d) at different time steps. Square marker represents fully decoherent regime, solid line the intermediate regime and dashed line indicates a fully coherent state.}
\label{fig:prof}
\end{figure}

\begin{figure}
\centering
\includegraphics[width=0.55\columnwidth]{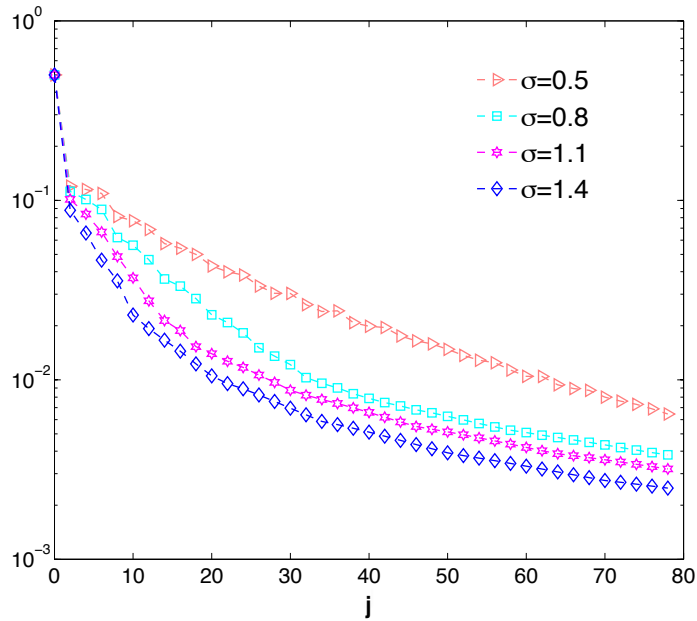}
\caption{Log-lin plot of time evolution of the spin coherence $C_{j}$ in a random $\theta$-field
%VS time step $j$ 
for various values of the noise parameter $\sigma$.   
%Grid points $n = 2^7$, $\Delta x$ = $2 \pi/n$. The symmetrical initial condition $\Psi_{j = 0, %m = n/2+1} = (b_L + i b_R)/\sqrt{2}$ and $\Psi_{j = 0, m} = 0$ if $m \ne n/2+1$. 
}
\label{fig:coe}
\end{figure}

\begin{figure}
\centering
\includegraphics[width=0.85\columnwidth]{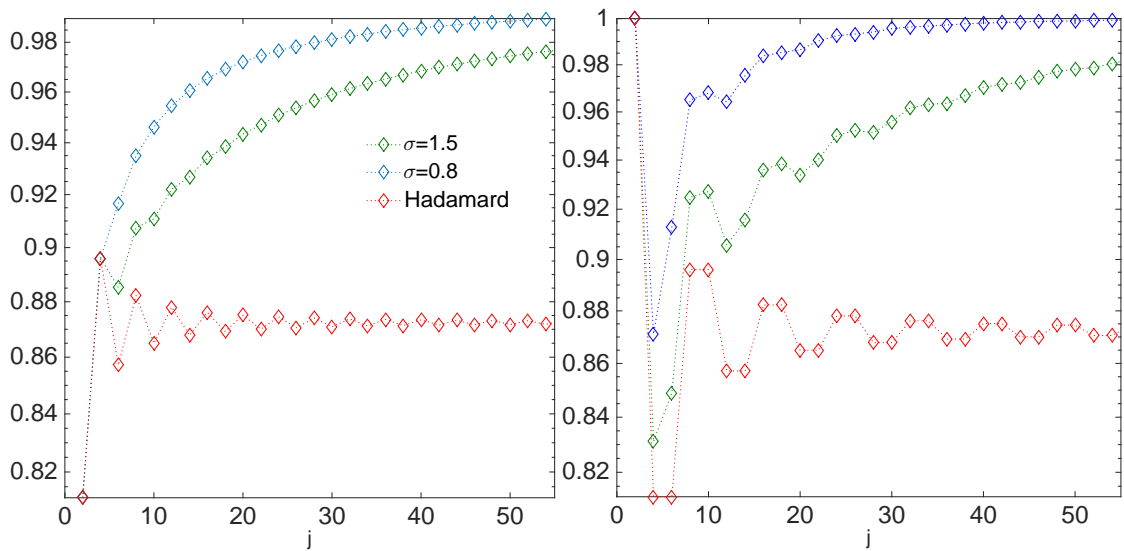}
\caption{Log-lin plot of time evolution of Shannon entanglement entropy $S_r$ compared to the Shannon entanglement entropy of the average transport in a random $\theta$-field (left) and $\xi$-field (right)  }
\label{fig:entr}
\end{figure}

%\subsection{Density profiles and decoherence}

Typical density profiles of the average transport are shown in Figure \ref{fig:prof} for random gravitational and random electric fields. 
%These and all subsequent numerical results have been obtained by direct simulation of the dynamics defined by ${\bar {\mathcal R}}^{g}$. 
For small enough values of the noise parameter $\sigma$, the average transport behaves at short times like the Hadamard walk and is ballistic. Ballistic behavior then gradually disappears and is replaced by diffusive behavior. For larger values of $\sigma$, ballistic behavior is replaced, even at short times, by diffusive behavior. Note that the Gaussian-like form of the asympotic density profiles presents a central dip when the DTQWs interact with random gravitational fields, but presents a central cusp when the DTQWs interact with random electric fields.

Asymptotically, DTQWs in electric and gravitational fields which are random in time thus behave like classical random walks. This means that the randomness in the fields prompts the DTQWs to loose coherence. This can be confirmed by considering the spin coherence
%, which is 
defined by
\begin{equation}
C_j = \mbox{max}_{m, m'} \mid \rho^{LR}_{j, m, m'} \mid.
%= \mbox{max}_{m, m'} \mid \rho^{RL}_{j, m, m'} \mid.
\end{equation}
%The proof of equivalence comes from the symmetry of the initial condition and the symmetry of the operator (12) and (13). Infact $\rho^{RL}_{j = 0,K, p}$ under conjugation gives $\rho^{LR}_{j = 0,K,p}$ and the evolution operator doesn't change it. If we take the complex conjugation of $\rho^{RL}_{j = 1, K, p}$ we obtain $\rho^{LR}_{j = 1, -K, p}$, but the operator (12) and (13) keep the evolution symmetric around $K$=0 such that  $\rho^{LR}_{j = 1, -K, p}$=$\rho^{LR}_{j = 1, K, p}$. The relation (16) is verified because the fourier transform keep unchanged these considerations. 
Figure \ref{fig:coe} displays the typical time-evolution of the spin coherence for various values of the noise parameter $\sigma$. These results confirm that the average transport loses spin coherence and that a higher value of $\sigma$ leads to a quicker loss of spin coherence.

A brief comment on spatial coherence is in order. The retained initial condition vanishes everywhere except at $m = m' = 0$. If one prefers, the Fourier transform of the initial density operator is flat in both $K$ and $p$ space. There is thus initially no spatial coherence. As time increases, the Fourier transform ${\hat \rho} (K, p)$ of the density operator $\rho$ becomes non flat in both $K$ and $p$ (see for example the asymptotic form (\ref{eq:hatrhoNe4M}) of $\hat \rho$). In other words, each $K$-mode acquires spatial coherence. But ${\tilde \rho} (p) = \sum_K {\hat \rho} (K, p)$ remains flat in $p$ (data not shown) {\sl i.e.} there is no {\sl total} gain of spatial coherence. 

%\subsection{Entropy measurement}

The entanglement of the averaged dynamics can also be quantified by the Shannon entropy $S_r$ of the reduced density operator $\rho_r$ in spin space. To be precise \cite{kollar14, liu10, chandra10a}, 
$\rho_{r} = \sum_m{\rho_{m,m'= m}}$ and the Shannon entropy $S_r = - \mbox{tr}(\rho_r\log(\rho_r))$.
The time-evolution of $S_r$ is prensented is Figure \ref{fig:entr}, together with the entanglement entropy of the pure Hadamard walk with the same initial condition, which admits 
0.872 as asymptotic value \cite{abal06}. The increase in $S_r$ signals the loss of coherence and the figure confirms that this  decoherence by noise gets more effective as $\sigma$ increases. 

The scaling of the decoherence time for small values of $\sigma$ can be evaluated by the following reasonning. As previously explained, the operator ${\bar {\mathcal R}}^{e/g}$ completely controls the average dynamics. For $\sigma = 0$, there is no noise and the DTQW never decoheres {\sl i.e.} the decoherence time is infinite. The first non-vanishing terms in the expansion of ${\bar {\mathcal R}}^{e/g}$ around $\sigma = 0$ are of second order in $\sigma$. Thus, per time step, the effect of the noise on the DTQW is of order $\sigma^2$ for small enough values of $\sigma$. The typical decoherence time therefore scales as $\sigma^{-2}$ for small values of $\sigma$.

The next section, together with the appendices, provides an analytical investigation of how coherence is lost. In particular, the asymptotic form of the density operator is computed exactly. The corresponding density is Gaussian, which confirms that the DTQW behaves asymptotically like a classical random walk. Also, the asymptotic density operator is proportionnal to $u_1 = v_1 + v_4 = b_L \otimes b_L + b_R \otimes b_R$. This proves that the spin coherence, which measures the amplitude of the $b_L \otimes b_R$ component, vanishes asymptotically, in accordance with Figure \ref{fig:coe}.

\section{Quantitative description of the asymptotic regime}

\subsection{Central limit theorem}

The average dynamics is entirely determined by the eigenvalues $\lambda^{e/g}_r$ and corresponding eigenvectors $w^{e/g}_r$, $r = 1, 2, 3, 4$, of the operators ${\bar {\mathcal R}}^{e/g}$. As evident from Figure \ref{fig:prof}, the density profiles of the average transport become larger and smoother with time. This suggest that the asymptotic dynamics can be understood by computing the eigenvalues and eigenvectors only for values of $K$ much smaller than unity. The detailled analysis, though very instructive, is too involved to merit inclusion in the main body of this article and it is therefore presented in the Appendix. The main conclusion can be stated as follows. 

\newtheorem*{Theo}{Theorem}

\begin{Theo}

Let $K_j= K_*/\sqrt{j}$ where $K_*$ is an arbitrary but $j$-independent wave number. 
The average density operator in Fourier space admits as the time $j$ tends to infinity
%, for wave numbers of the form $K_j= K_*/\sqrt{j}$ where $K_*$ is an arbitrary but $j$-independent wave number, 
the following approximate asymptotic expression:
%\begin{equation}
%{\hat \rho}^{e/g}_{j = J} (K, p) =  \frac{1}{2} 
%\left( 1 - \alpha^{e/g} (p, \sigma) K^2 \right)^J u_1. 
%\label{eq:hatrhoNe3}
% \end{equation}
%In particular, for $K_J= K_*/\sqrt{J}$ (where $K_*$ is an arbitrary but $J$-independent wave number) and large enough $J$,
\begin{equation}
{\hat \rho}^{e/g}_{j} (K_j, p) 
%=  \frac{1}{2} 
%\left( 1 - \alpha^{e/g} (p, \sigma) \frac{K_*^2}{J} \right)^J u_1 
\sim  
%\frac{1}{2} \exp \left( - \alpha^{e/g} (p, \sigma) K_*^2\right) u_1
%=  
\frac{1}{2} \exp \left( - \alpha^{e/g} (p, \sigma) j K_j^2\right) u_1
\label{eq:hatrhoNe4M}
\end{equation}
where
\begin{equation}
\alpha^{e}(p, \sigma) =  2\,  \frac{3 + \left(\text{sinc}(\sigma) \right)^2 + 2 \left(\text{sinc}(\sigma/2) \right)^2
\left( 1 + \text{sinc}(\sigma) \right) + 4 \cos(2p) \left( \text{sinc}(\sigma)  + \left(\text{sinc}(\sigma/2) \right)^2 \right)
}{
3 + \left(\text{sinc}(\sigma) \right)^2 - 2 \left(\text{sinc}(\sigma/2) \right)^2
\left( 1 + \text{sinc}(\sigma) \right) + 4 \cos(2p) \left( \text{sinc}(\sigma)  - \left(\text{sinc}(\sigma/2) \right)^2 \right)
}
\label{eq:alphae}
\end{equation}
and 
\begin{equation}
\alpha^{g}(p, \sigma) =  2 \, \frac{1 + \left(\text{sinc}(\sigma)\right)^2}{1 - \left(\text{sinc}(\sigma)\right)^2}.
\label{eq:alphag}
\end{equation}
\end{Theo}

This result is a central limit theorem which proves that the asymptotic density operator is approximately Gaussian in $K$-space, with a typical width (in $K$-space) which decreases as $j^{-1/2}$, as in classical random walks and non quantum diffusions. Note that $\alpha^g$ is actually independent of $p$. 

One of the consequences of (\ref{eq:hatrhoNe4M}) is that the projection of ${\hat \rho}_{j = J} (K_J, p)$ on the subspace spanned by $(u_2, u_3, u_4)$ tends to zero. Remembering the expressions of the $u_i$ in termes of $b_L$ and $b_R$, this means that ${\hat \rho}^{LR}$, ${\hat \rho}^{LR}$ and ${\hat \rho}^{LL} - {\hat \rho}^{RR}$ all tend to zero as $J$ tends to infinity. The component along $u_1$ coincides with ${\hat \rho}^{LL} + {\hat \rho}^{RR}$ and determines the asymtotic density of the averaged walk after summation over $p$ and Fourier transform over $K$.

%Since $K_*$ is arbitrary, this last approximate expression for $\hat \rho$ is valid for all spatial scales $1/K$ which vary as $\sqrt J$ {\sl i.e.} for all spatial scales where diffusion is expected to occur. 

\subsection{Asymptotic mean-square displacement}

Let us now explicitly compute the asymptotic expression of the mean-square displacement ${\overline {m^2}}^{\, e/g}$ in the special case of a random DTQW on the infinite line. Switching back the original spatial variables $m$ and $m'$ involves a double integration over $K$ and $p$. The $2D$ measure to be used in this integration is $dk dk' = 2 dK dp$. The density $N^{e/g}_{j m}$ at time $j$ and point $m$ is the trace over $m' = m$ of the component of the density operator along the basis vector $u_1$.
Expression (\ref{eq:hatrhoNe4}) for ${\hat \rho}^{e/g}$ is only valid for $K \ll 1$ (see the Appendix). But the functions $\alpha^{e/g}(p, \sigma)$ are always non vanishing. The width $\Delta K(j, p)$ of $\hat \rho^{e/g}_{j} (K, p)$ in $K$ thus scales as $1/\sqrt{j}$ and tends to zero as $j$ tends to infinity. Thus, for large enough $j$, the density and mean square displacement are given by:
\begin{equation}
N^{e/g}_{j m} = \frac{1}{4 \pi^2}\, \int_{p = - \pi}^{\pi} dp \int_{K = - \pi}^{\pi} dK \exp \left( - \alpha^{e/g} (p, \sigma) j K^2\right) \exp \left( - i K m \right)
\label{eq:defNegjm}
\end{equation}
and
\begin{equation}
{\overline {m^2}}^{\, e/g} (j, \sigma) = \frac{1}{4 \pi^2}\, \sum_{m \in \mathbb Z} m^ 2 \int_{p = - \pi}^{\pi} dp \int_{K = - \pi}^{\pi} dK  \exp \left( - \alpha^{e/g} (p, \sigma) j K^2\right) \exp \left( - i K m \right).
\label{eq:defbarm2}
\end{equation}
Since the width $\Delta K(j, p)$ of $\hat \rho_{j, K, p}$ scales as $1/\sqrt{j}$, one can also replace all discrete summations over $m$ by integrals over the real line, because $\Delta K(j, p) \times \Delta x = 1/\sqrt{j} \times 1 \ll 1$ for large enough $j$. Indeed, a simple computation confirms that the integrated density $\int_{\mathbb R} dm N^{e/g}_{jm}$ (with $N^{e/g}_{jm}$ given by (\ref{eq:defNegjm})) is equal to unity at all times $j$. Replacing in (\ref{eq:defbarm2}) the discrete summation over $m$ by an integral delivers
\begin{equation}
{\overline {m^2}}^{\, e/g}(j, \sigma)   =  \frac{j}{\pi} \int_{- \pi}^{\pi} \alpha^{e/g} (p, \sigma) dp.
\end{equation}
The computation of ${\overline {m^2}}^{\, g} (j, \sigma)$ is trivial because $\alpha^{g} (p, \sigma)$ does not depend on $p$. One finds
%\begin{equation}
%{\overline {m^2}}^{\, e} (J, \sigma) =  2J \, \frac{\text{sinc}(\sigma)  + \left(\text{sinc}(\sigma/2) \right)^2}{ \text{sinc}(\sigma)  - \left(\text{sinc}(\sigma/2) \right)^2 } 
%\end{equation}
%and
%\begin{figure}
%\centering
%\includegraphics[width=0.40\columnwidth]{alphaC.pdf}
%\includegraphics[width=0.40\columnwidth]{thetaC.pdf}
%\includegraphics[width=0.40\columnwidth]{coefA.pdf}
%\includegraphics[width=0.40\columnwidth]{coefNA.pdf}
%\caption{Asymptotic diffusion coefficients of the avarage transport $D^{e}(\sigma)$ (solid, red) VS $\sigma$ with $\sigma \leq \pi$ and  $D^{g}(\sigma)$ (dashed, blue) with $\sigma \leq \pi$. $D^{e}(\sigma)$ has a discontinuity for $\sigma$= 2 $\textit{s}$ $\pi$, with $\textit{s} \in \mathbb{Z}$.}
%\label{fig:coeff}
%\end{figure}

\begin{equation}
{\overline {m^2}}^{\, g} (j, \sigma) =  2 D^g(\sigma) j
\end{equation}
with
\begin{equation}
D^g(\sigma) = 
2 \,\frac{1 + \left(\text{sinc}(\sigma)\right)^2}{1 - \left(\text{sinc}(\sigma)\right)^2}.
\label{eq:Dg}
\end{equation}
The exact expression for ${\overline {m^2}}^{\, e} (j, \sigma)$ is more involved. A direct computation leads to:
\begin{equation}
{\overline {m^2}}^{\, e} (j, \sigma) =  2 D^e (\sigma) j
\end{equation}
with
\begin{equation}
D^e(\sigma) = \, \frac{2}{\text{sinc}(\sigma) - \left(\text{sinc}(\sigma/2)\right)^2}\, 
\left(\left( \text{sinc}(\sigma) + \left(\text{sinc}(\sigma/2)\right)^2\right)
+ \frac{2 \left(\text{sinc}(\sigma/2)\right)^2 \left(
\left(\text{sinc}(\sigma)\right)^2 + 2 \text{sinc}(\sigma) - 3
\right)}
{s(\sigma)}
\right)
\label{eq:De}
\end{equation}
with
\begin{equation}
s(\sigma) = \left[\left(3 + \left(\text{sinc}(\sigma)\right)^2 - 2 \left(\text{sinc}(\sigma/2)\right)^2
\left(1 + \text{sinc}(\sigma)
\right)\right)^2 - 16 \left(\text{sinc}(\sigma) - \left(\text{sinc}(\sigma/2)\right)^2\right)^2
\right]^{1/2}.
\end{equation}

In both electric and gravitational case, the asymptotic mean square displacement in physical space grows linearly in time, as for classical random walks and non quantum diffusions. The functions $D^e$ and $D^g$ are the asymptotic diffusion coefficients of the average transport. Both functions are strictly decreasing on $(0, 2 \pi)$. Thus, decoherence occurs more rapidly as $\sigma$ increases (see Section \ref{sec;Qualit}), but the asymptotic diffusion coefficients decrease with $\sigma$. We also note that $D^g (\sigma) < D^e(\sigma)$ for all $\sigma \in (0, 2 \pi)$.

\begin{figure}
\centering
\includegraphics[width=0.40\columnwidth]{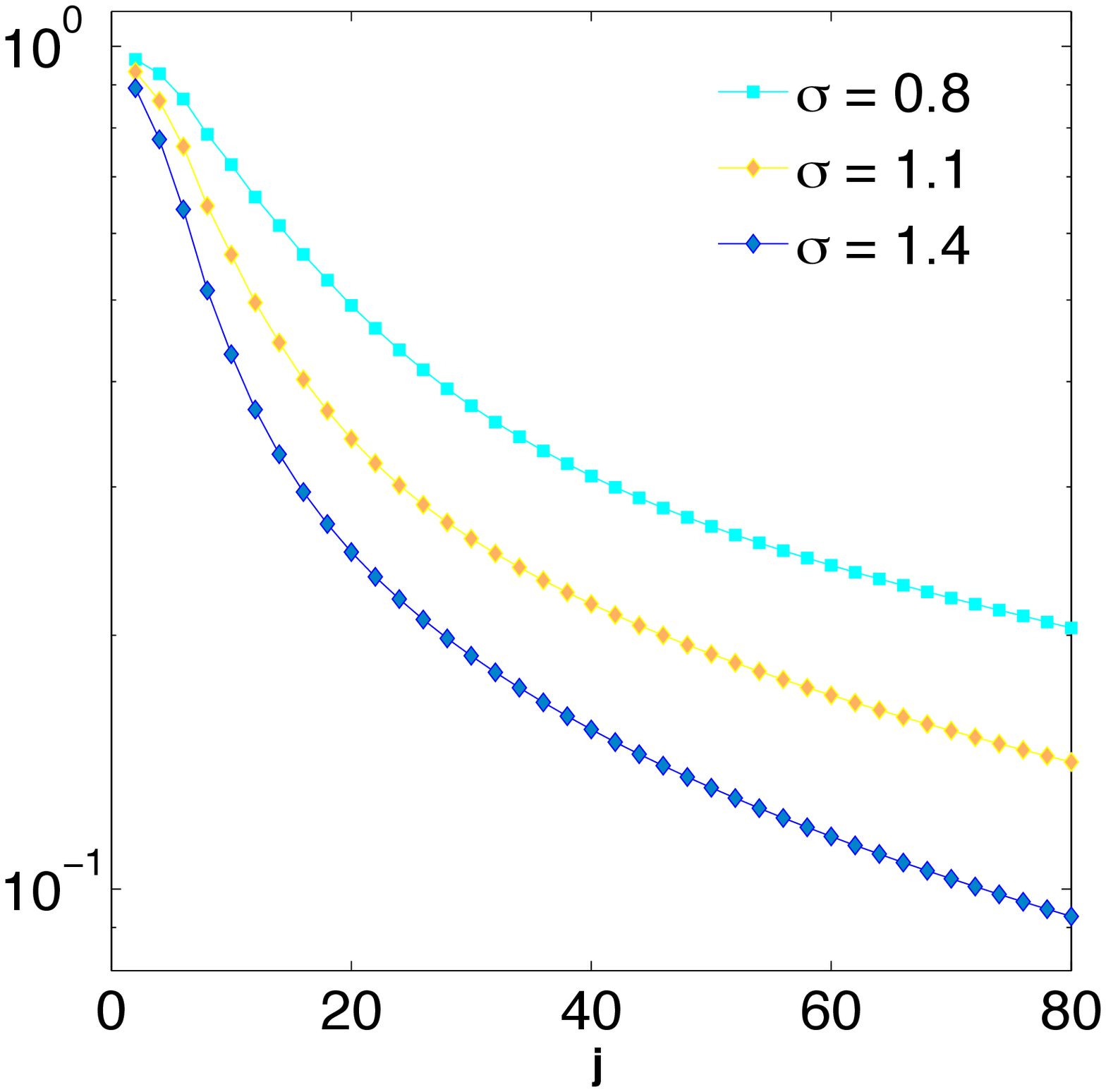}
\includegraphics[width=0.40\columnwidth]{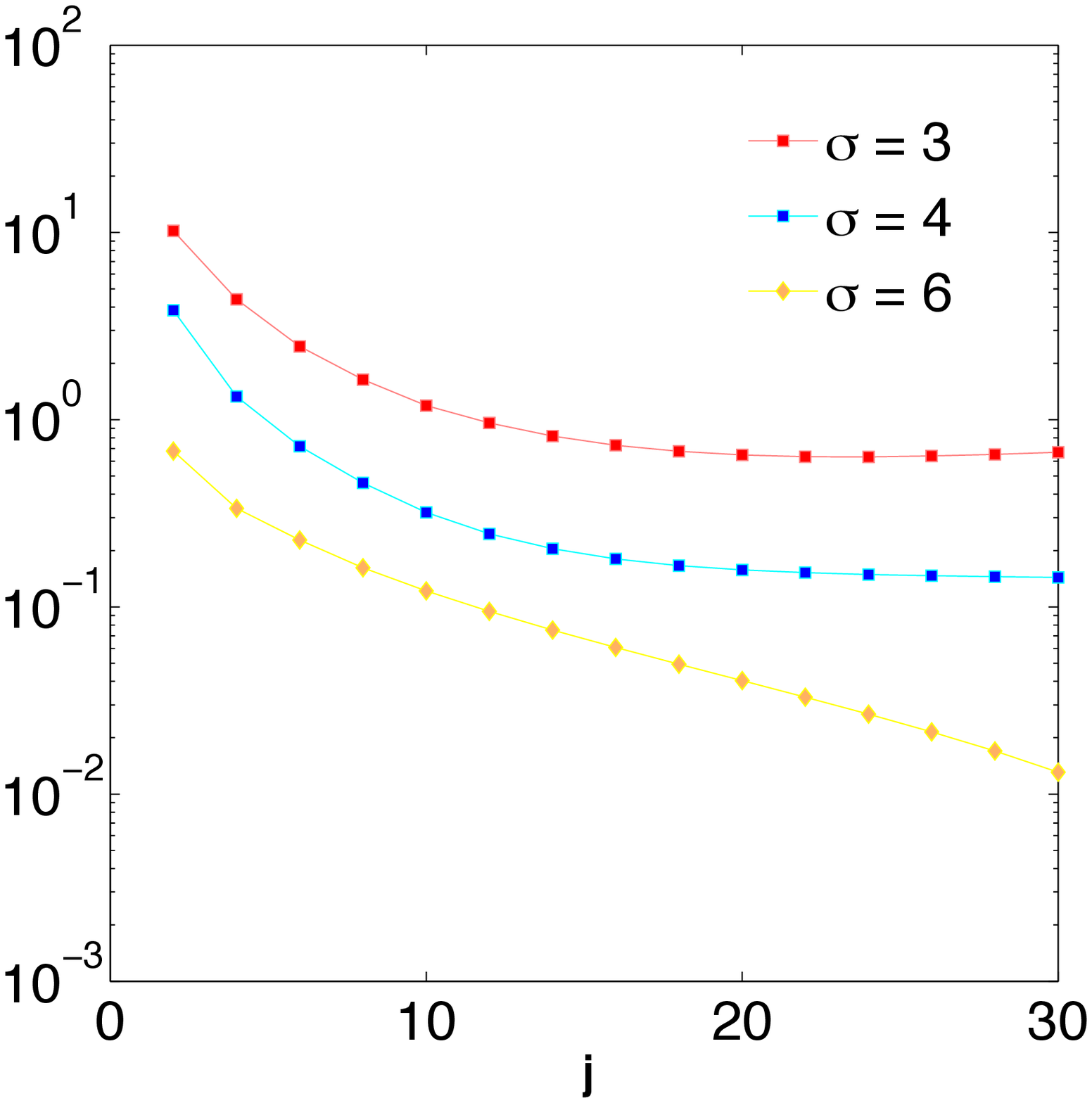}
\caption{Left (right) figure: time-evolution of the relative difference between the gravitational (eletric) diffusion coefficients computed from numerical simulations and the exact analytical expressions. 
%from the mean square displacement. It has been computed numerically for various values of $\sigma$. The relative difference is ($\overline{m^2}^{\, e,g} (J, \sigma)$ -$2 D^{e/g} (\sigma) J$)/$2 D^{e/g} (\sigma) J$ where $\overline{m^2}^{\, e,g} (J, \sigma)$ computed numerically $\sum_m m^2 N(j,m)$.
}
\label{fig:diff}
\end{figure}

Figure \ref{fig:diff} shows the time-evolution of the relative difference between the  diffusion coefficients computed from (\ref{eq:Dg}), (\ref{eq:Dg}) and the mean square displacement computed from numerical simulations for various values of $\sigma$. 
%Similar results are obtained for
%${\overline m^2}^{\, e} (J, \sigma)$ (see Figure \ref{fig:diff}.b ). 
This figures clearly supports the analytical computation presented in this Section.
 
%Also, equivalents for $\sigma<<1$. 
%Apparently, $D^g \sim a/\sigma^2$ while $D^e \sim b/\sigma^3$.

\section{Continuous limit}
\label{sec:CL}

The formal continuous limit of the original, unaveraged evolution equations (\ref{eq:defdens}) and (\ref{eq:defwalkdiscr}) has already been considered in \cite{DMD13b,DMD14} and coincides with the Dirac equation obeyed by a fermion minimally coupled to an electric field and/or a relativistic gravitational field. Let us now determine the formal continuous limit of the averaged evolution equations specified by the operators ${\bar {\mathcal R}}^{e}$ and ${\bar {\mathcal R}}^{g}$.

%\subsection{Electric case}

As shown and discussed in \cite{DMD13b,DMD14} for the unaveraged evolution equations, the object which admits a continuous limit for $\theta = \theta_H$ or $\xi = \xi_H$ is not the original walk, but the walk derived from it by keeping only one time step out of two \footnote{Note however that it is possible to keep all time steps if one is only interested in the continuous limit of the probability of the original walk}.
We thus search for the continuous limit of the following discrete equations:
\begin{equation}
{\hat \rho}_{j+2}(K, p) = \left({\bar {\mathcal R}}^{e/g}(K, p, \sigma)\right)^2 {\hat \rho}_{j}(K, p).
\end{equation}
To be specific, we restrain $j$ to uneven positive integer values and decide to work on the inifinite line, so that $K$ and $p$ take all values in $(-\pi, + \pi)$.

We now supppose that, for all  uneven $j = 2r + 1$,  $\rho_{j = 2r+1, m, m'}$ ({\sl resp.} ${\hat \rho}_{j = 2r +1} (K, p)$) is the value taken by a certain function $\rho$
({\sl resp.} ${\hat \rho}$) at `time' $t_r = r$ and positions $x_m = m$ and $x_{m'} = m'$ ({\sl resp.} momenta $K$ and $p$). 
Roughly speaking, the continuous limit refers to situations where the function $\rho$ ({\sl resp.} $\hat \rho$) varies only little during one time step 
%{\sl i.e.} has characteristic temporal variation scales much larger than the time-step 
$t_{r+1} - t_r = 1$. A necessary and sufficient condition for this to be realized is that $\left({\bar {\mathcal R}}^{e/g}(K, p, \sigma)\right)^2$ be close to unity. Direct inspection reveals that this transcribes into $\sigma \ll 1$, $K \ll 1$ and $p \ll 1$. The last two conditions mean that $\rho$ has caracteristic spatial variation scales much larger than the distance $m+1 - m = 1$ between adjacent grid points and the first condition states that the noise amplitude is small. Note that $K$, $p$ and $\sigma$ are {\sl a priori} independent inifnitesimal quantities. In particular, there is no reason why $K$ and $p$ should be of the same order of magnitude.

The formal continuous limit is then obtained by expanding $\left({\bar {\mathcal R}}^{e/g}(K, p, \sigma)\right)^2$ around $K = 0$, $p = 0$, $\sigma = 0$ and by replacing ${\hat \rho}_{j+2} - {\hat \rho}_j$ by $\partial_t {\hat \rho}$.
One thus gets equations of the form :
\begin{equation}
\partial_t {\hat \rho} (t, K, p) = \left( {\mathcal S}^{e/g}(K, p, \sigma) - 1 \right){\hat \rho} (t, K, p)
\end{equation}
where, for example, 
\begin{equation}
{\mathcal S}^g(K, p, \sigma) = 
\begin{bmatrix}
1 - 4K^2& 2i K & 2i K (1 - \frac{p^2}{2})(1 - \frac{\sigma^2}{6})  & 2K p  (1 - \frac{\sigma^2}{6})\\
2i K (1 - \frac{p^2}{2})(1 - \frac{\sigma^2}{3}) & (1 - 4K^2) (1 - \frac{p^2}{2})(1 - \frac{\sigma^2}{3}) & p^2 (1 - \frac{\sigma^2}{6}) & 2ip (1 - \frac{\sigma^2}{6}) \\
2 i  K (1 - \frac{\sigma^2}{6}) &  - 4 K^2  (1 - \frac{\sigma^2}{6})  & (1 - 4K^2) (1 - \frac{p^2}{2})(1 - \frac{\sigma^2}{3}) & - i p (1 - 4K^2) (1 - \frac{\sigma^2}{3}) \\
-2 K p  (1 - \frac{\sigma^2}{6}) & i p (1 - 4K^2) (1 - \frac{\sigma^2}{6})   & - i p  & 1 - 4 p^2
\end{bmatrix}
\label{eq:defS}
\end{equation}
at second order in all three independent infinitesimals $K$, $p$ and $\sigma$. These equations can be translated into physical space by remembering that $-iK$ and $-ip$ are the Fourier representations of $\partial_X$ and $\partial_y$ where $X = (x + x')/2$ and $y = x' - x$. 

The analysis presented in Sections III and IV above has been carried out with an initial condition which spreads over the whole $K$- and $p$-ranges. The resulting density operator does localize in time around $K = 0$, but it never localizes around $p = 0$ and remains spread in $p$-space. The continuous limit thus cannot be used to recover the results of Section IV. As can be checked directly from (\ref{eq:defS}), the continuous limit equations nevertheless predict diffusive behavior if $K$ is much lower than both $p$ and $\sigma$. A systematic study of the continuous limit dynamics for various scaling laws obeyed by $K$, $p$ and $\sigma$ falls outside the scope of this article and will be presented elsewhere.

%%%%%%%%%%%%%%%%%%%%%%%%%%%%%%%%%%%%%%%%%%%
\section{Conclusion}

We have studied two families of DTQWs which can be considered as simple models of quantum transport of a Dirac fermion in random electric or gravitational fields. We have
proven analytically and confirmed numerically that randomness of the fields in time leads on average to decoherence of the walks. The asymptotic average transport is thus diffusive and we have computed exactly the diffusion coefficients. We have also obtained and discussed the continuous limit of the model. 

A few words about the loss of coherence in DTQWs may prove useful at this point. Pure, deterministic DTQWs are standard quantum systems in the sense that their time-evolution is unitary. They thus never loose coherence nor do they exhibit diffusive behavior. As with any quantum system, the loss of coherence in DTQWs is induced by the so-called interaction with an environment. There are essentially two ways to model this interaction. The first one is to start from the unitary evolution of the density operator and to modify this unitary evolution into a non-unitary one by introducing so-called projector or measurement operators \cite{Ken07a, brun2003, vieira14, chisaki2011}. The second way of introducing decoherence is the one followed in this article. It consists in introducing some randomness in the parameters of the DTQW and in averaging over this randomness \cite{Werner11, Cedzich12, Joye11,perez10, obuse2011topological}. Contrary to the unaveraged density operator, the averaged density operator then follows a non-unitary evolution and this breakdown of unitarity induces the loss of coherence and the asymptotic diffusive behavior displayed by the averaged transport. 

The results of this article constitute/are an addition to the already extensive literature dealing with the asymptotic behavior of DTQWs and CTQWs. Standard deterministic QWs are famous for typically exhibiting asymptotic ballistic behavior. But diffusive and anomalous diffusive asymptotic behavior have also been observed \cite{shikano2014discrete, prokof2006decoherence}, as well as localization \cite{obuse2011topological, schreiber2011decoherence, inui2004localization} and soliton-like structures \cite{perez10}. 

Let us conclude by listing a few natural extensions of this work. The random artificial gauge fields considered in this article have two main characteristics: they depend only on time and the associated mean fields vanish \cite{D04a}. One should therefore extend the analysis presented above to situations where the mean fields do not vanish and where the artificial gauge fields depend not only on time, but also on position. In particular, the continuous limit equation derived in Section \ref{sec:CL} is markedly different from both the Caldeira-Leggett \cite{CaldLegg87,Cald83} and the relativistic Kolmogorov equation describing relativistic stochastic processes \cite{DMR97a,CD07g,DEF12a}. Indeed, because the random fields depend only on time, the dynamics considered in this article does not couple different $(K, p)$-modes, but these are coupled in both the Caldeira-Leggett and the relativistic Kolmogorov equation. Considering DTQWs coupled to artificial gauge fields which also depend randomly on position should therefore lead to master equations closer to the the Caldeira-Legget and the Kolmogorov models. Moreover, cases where both electric and gravitational fields vary randomly are certainly worth investigating. 

Finally, at least some DTQWs in two spatial dimensions can be considered as models of quantum transport in electromagnetic fields \cite{DMD13a}. 
The analysis presented in this article should therefore be repeated in higher dimensions to include random magnetic fields \cite{DA14} and evaluate their effects on spintronics.
%%%%%%%%%%%%%%%%%%%%%%%%%%%%%%%%%%%%%%%%%%%%%%%%

%\section*{Appendix}

\appendix

\section{Interpretation in terms of artificial gauge fields}

It has been proven in \cite{DMD12a, DMD13b,DMD14} that quantum walks in $(1 + 1)$ dimensional space-times can be viewed as modeling the transport of a Dirac fermion in artificial electric and gravitational fields generated by the time-dependance of the angles $\theta$ and $\xi$. We recall here some basic conclusions obtained in \cite{DMD12a, DMD13b,DMD14} and also offer new developments useful in interpreting the results of the present article.

The DTQWs defined by (\ref{eq:defwalkdiscr}) are part of a larger family whose dynamics reads:
\begin{equation}
\begin{bmatrix} \psi^{L}_{j+1, m }\\ \psi^{R}_{j+1, m} \end{bmatrix} \  = 
{\tilde {\mathcal B}}\left(\theta_{j, m} ,\xi_{j, m}, \zeta_{j, m}, \alpha_{j, m} \right)
 \begin{bmatrix} \psi^{L}_{j, m+1} \\ \psi^{R}_{j, m-1} \end{bmatrix},
\label{eq:defwalkdiscrtilde}
\end{equation}
where 
\begin{equation}
 {\tilde {\mathcal B}}(\theta ,\xi, \zeta, \alpha) = e^{i \alpha}
\begin{bmatrix}  e^{i\xi} \cos\theta &  e^{i \zeta} \sin\theta\\ - e^{i \zeta}  \sin\theta &  e^{-i\xi} \cos\theta
 \end{bmatrix}.
\label{eq:defBtilde}
\end{equation} 
The walks in this larger family are characterized by three time- and space-dependent Euler angles $(\theta, \xi, \zeta)$ and by a global, also time- and space-dependent phase $\alpha$. They have been shown to model the transport of Dirac fermions in artificial electric and relativistic gravitational fields generated by the time-dependence of the three Euler angles and of the global phase. In a $(1 + 1)$ dimensional space-time, an electric field derives from a $2$-potential $A_{j, m}  = \left(V_{j, m}, {\mathcal A}_{, mj} \right)$ and a relativistic gravitational field is represented by $2D$ metrics $G_{j, m}$. The walks considered in this article correspond to 
\begin{eqnarray}
\xi & = & \frac{\pi}{2} + {\bar \xi}_j \nonumber \\
\theta & = & \frac{\pi}{4} + {\bar \theta}_j  \nonumber \\
\alpha & = & \frac{\pi}{2} + {\bar \alpha} \nonumber \\
\zeta & = & 0
\end{eqnarray}
where ${\bar \xi}_j$ and ${\bar \theta}_j  $ are random variables which depend on the time $j$ and ${\bar \alpha} = 3 \pi/2$. According to \cite{DMD14}, these walks model the transport of a Dirac fermion in an electric field generated by the 2-potential 
\begin{equation}
A_{j}  = \left(V_{j}, {\mathcal A}_{j} \right) = \left( {\bar \alpha}_{j}, - {\bar \xi}_{j}\right) = \left(\pi/2, - {\bar \xi}_{j}\right) 
\end{equation}
and in a gravitational field caracterized by the metric
\begin{equation}
G_{j} = \mbox{diag} \left(1, - \cos^{-2} (\theta_j)\right).
\label{eq:metricG}
\end{equation}

Since relativistic gravitational fields are represented by space-time metrics \cite{W84a}, making the angle $\theta$ a time-dependent random variable is equivalent to imposing a time-dependent random gravitational field. 
To better understand the electric aspects of the problem, let us recall that the DTQWs defined by (\ref{eq:defwalkdiscrtilde}) exhibit the following exact discrete gauge invariance \cite{DMD14}:
\begin{eqnarray}
\Psi'_{j, m} & = & \Psi_{j, m} e^{- i \phi_{j, m}} \nonumber \\
\xi'_{j, m} & = & \xi_{j, m} + \delta_{j, m} \nonumber \\
\theta'_{j, m} & = & \theta_{j, m} \nonumber \\
{\alpha}'_{j, m} & = & \alpha_{j, m} + \frac{\sigma_{j, m}}{2} \nonumber \\
\zeta'_{j, m} = & = & \zeta_{j, m} - \delta_{j, m}
\end{eqnarray}
where 
\begin{eqnarray}
\sigma_{j, m} & = & \phi_{j, m+1} + \phi_{j, m-1} - 2 \phi_{j +1, m} \nonumber \\
\delta_{j, m} & = & \frac{\phi_{j, m+1} - \phi_{j, m-1} }{2}
\end{eqnarray}
and $\phi$ is an arbitrary time- and space-dependent phase shift. Let us now define a new quantity $E_{j, m}$ by
\begin{equation}
E_{j, m} = - \left({\mathcal D}_s V\right)_{j, m} + \left( {\mathcal D}_t {\mathcal A}\right)_{j, m}
\end{equation}
where
%\begin{equation}
%{\mathcal D}_t = {\mathcal T} - {\mathcal S}
%\end{equation}
the actions of the operators ${\mathcal D}_s$ and ${\mathcal D}_s$ on an arbitrary time- and space-dependent quantity $u_{j, m}$ are
\begin{equation}
\left({\mathcal D}_s u\right)_{j, m} = \frac{u_{j, m+1} - u_{j, m-1}}{2}
\end{equation}
and
\begin{equation}
\left({\mathcal D}_t u\right)_{j, m} = \frac{2 u_{j+1, m} - u_{j, m+1} - u_{j, m-1}}{2}.
\end{equation}
The operators ${\mathcal D}_s$ and ${\mathcal D}_t$ are discrete counterparts of space- and time-derivatives. It is straightforward to check that the quantity $E_{j, m}$ is gauge invariant and coincides, in the continuous limit, with the standard electric field $E(t, x)$, defined by $E(t, x) = - \partial_x V + \partial_t {\mathcal A}$. The quantity $E_{j. m }$ is thus a {\sl bona fide} electric field in discrete space-time. For the DTQWs considered in this article, this electric field depends only on the time $j$ and is related to the angle ${\bar \xi}$ by
$E_{j} = - \left( {\bar \xi}_{j+1} - {\bar \xi}_j\right)$.
Making this angle a time-dependent random variable is thus equivalent to imposing a random electric field.

%\section{Detailled analysis of the asymtotic dynamics}

\section{Aymptotic computation of the eigenvalues and eigenvectors of the averaged transport operators}

%The average dynamics is entirely determined by the eigenvalues $\lambda^{e/g}_r$ and corresponding eigenvectors $w^{e/g}_r$, $r = 1, 2, 3, 4$, of the operators ${\bar {\mathcal R}}^{e/g}$. As evident from Figure \ref{fig:prof}, the density profiles of the average transport become larger and smoother with time. This suggest that the asymptotic dynamics can be understood by computing the eigenvalues and eigenvectors only for values of $K$ much smaller than unity. 

Let us here compute the eigenvalues $\lambda^{e/g}_r$ and eigenvectors $w^{e/g}_r$, $r = 1, 2, 3, 4$ only for values of $K$ much smaller than unity. We do not perform an expansion in $p$ because the initial condition is uniform in $p$ and the average evolution does not localize the density operator around $p = 0$. Indeed, the initial condition is localized at $x' = x$ {\sl i.e.} does not exhibit any spatial correlation and the dynamics does not create spatial correlations.

%All expansions below are valid for finite values of $p$. They are thus valid only if $\mid K \mid \ll \mid p \mid$. This does not invalidate the asymptotic computation, at least on the infinite line, because, as time increases, the density operator becomes more and more localized around $K = 0$. For simplicity purposes, the analytical discussion will thus be restricted to DTQWs on the infinite line.

The second order expansions of the operators ${\bar {\mathcal R}}^e$ and ${\bar {\mathcal R}}^g$ in $K$ read:
\begin{equation}
\displaystyle{
{\bar {\mathcal R}}^e_2(K, p, \sigma) = 
\begin{bmatrix}
1 - 2K^2 & 2iK & 0  & 0 \\
0 & 0  & \text{sinc}(\sigma/2) \cos (2p)& -  i\,  \text{sinc}(\sigma/2) \sin (2p) \\
2i \text{sinc}(\sigma/2) K &  \text{sinc}(\sigma/2) \left(1 - 2 K^2 \right)& \frac{ (1 -  \text{sinc}(\sigma))}{2} \cos (2p)& i \frac{( \text{sinc}(\sigma) - 1)}{2} \sin(2p)\\
0 & 0 & i  \frac{(1 +  \text{sinc}(\sigma)) }{2}\sin(2p) & - \frac{( \text{sinc}(\sigma) + 1)}{2}  \cos(2p)
\end{bmatrix},}
\end{equation}
and
\begin{equation}
{\bar {\mathcal R}}^g_2(K, p, \sigma) = 
\begin{bmatrix}
1 - 2K^2& 2i K & 0  & 0 \\
0 & 0  & \text{sinc}(\sigma) \cos (2p)& -  i\,  \text{sinc}(\sigma) \sin (2p) \\
2 i \text{sinc}(\sigma) K &  \text{sinc}(\sigma)   &  0 & 0\\
0 & 0 & i \sin(2p) & -  \cos(2p)
\end{bmatrix}.
\end{equation}

%At zeroth order in $K$, the operators ${\bar {\mathcal R}}_e$ and ${\bar {\mathcal R}}_g$ read
%%\begin{equation}
%%{\bar {\mathcal R}}^{e/g} (\sigma, K, p) =  {\bar {\mathcal R}}^{e/g}_0 (p, \sigma) +\Delta {\bar {\mathcal R}}^{e/g} (p, \sigma) + O(K^3)
%%\end{equation}
%%where
%\begin{equation}
%\displaystyle{
%{\bar {\mathcal R}}^e(K = 0, p, \sigma) = 
%\begin{bmatrix}
%1 & 0 & 0  & 0 \\
%0 & 0  & \text{sinc}(\sigma/2) \cos (p)& -  i\,  \text{sinc}(\sigma/2) \sin (p) \\
%0 &  \text{sinc}(\sigma/2) & \frac{ (1 -  \text{sinc}(\sigma))}{2} \cos (p)& i \frac{( \text{sinc}(\sigma) - 1)}{2} \sin(p)\\
%0 & 0 & i  \frac{(1 +  \text{sinc}(\sigma)) }{2}\sin(p) & - \frac{( \text{sinc}(\sigma) + 1)}{2}  \cos(p)
%\end{bmatrix},}
%\end{equation}
%and
%\begin{equation}
%{\bar {\mathcal R}}^g(K = 0, p, \sigma) = 
%\begin{bmatrix}
%1 & 0 & 0  & 0 \\
%0 & 0  & \text{sinc}(\sigma) \cos (p)& -  i\,  \text{sinc}(\sigma) \sin (p) \\
%0 &  \text{sinc}(\sigma)   &  0 & 0\\
%0 & 0 & i \sin(p) & -  \cos(p)
%\end{bmatrix}.
%\end{equation}
For $K = 0$, these two matrices are both block diagonal and we write ${\bar {\mathcal R}}^{e/g}_2(K = 0, p, \sigma) = \mbox{diag} (1, M^{e/g}(p, \sigma))$, where $M^{e/g}(p, \sigma)$ are $3 \times 3$ matrices acting in the space spanned by $(u_2, u_3, u_4)$. The matrices ${\bar {\mathcal R}}^{e/g}_2(K = 0, p, \sigma)$ share $u_1$ as common eigenvector, which we identify as $w^{e/g}_1(K = 0, p, \sigma)$; the associated eigenvalue is $\lambda_1^{e/g}(K = 0, p, \sigma) = 1$. The other eigenvectors and eigenvalues, at zeroth order in $K$, are those of $M^{e/g}(p, \sigma)$. These eigenvalues can be computed analytically by solving the third-order characteristic polynomials associated to these matrices. The explicit expressions of these eignevalues are quite involved and need not be replicated here. What is important is how the moduli of these eigenvalues compares to unity. Direct inspection reveals that the moduli of all three $\lambda^e_r(0, p, \sigma)$, $r = 2, 3, 4$ are strictly inferior to unity if $\sigma$ is not vanishing. The same goes for all three eigenvalues in the gravitational case, except for one of them which reaches $\pm 1$ independantly of $\sigma$ for $p = \pm \pi$ and is also equal to $+1$ for $p = 0$; the eigenspaces corresponding to $\lambda_4^g (\pm \pi, \sigma)$ and $\lambda_4^g (0, \sigma)$ are identical and generated by $u_4$, which we choose as $w_4^g(p = \pm \pi, \sigma) = w_4^g(0, \sigma)$. For other values of $p$, the eigenvalue $\lambda_4^g(p, \sigma)$ and the eigenvector $w_4^g(p, \sigma)$ are defined by continuity. All other eigenvectors need not be specified for what follows.

Let us now turn to non vanishing values of $K$. The characteristic polynomials of ${\bar {\mathcal R}}^{e/g}_2(K, p, \sigma)$ contain terms of order $2$ and $4$ in $K$; at lowest order in $K$, the corrections to the eigenvalues $\lambda_j^{e/g}(K = 0, p, \sigma)$ thus scale generically as $K^2$. Let $\lambda$ be the variable of the characteristic polynomials. At second order in $K$, the $K$-dependent correction to each of the zeroth order eigenvalues $\lambda_r^{e/g}(K = 0, p, \sigma)$ can be found by expanding the characteristic polynomial of ${\bar {\mathcal R}}^{e/g}_2(K, p, \sigma)$ at first order in $\left(\lambda - \lambda_r^{e/g}(K = 0, p, \sigma)\right)$ and by keeping only the terms scaling as $K^2$. This gives rational expressions for the corrections to the eigenvalues; these rational expressions can be further simpified by a final expansion around $K = 0$ if $p$ is treated as a finite, non infinitesimal quantity {\sl i.e.} $\mid K \mid \ll \mid p \mid$. One then finds:
\begin{equation}
\lambda_1^{e/g}(K, p, \sigma) = 1 - \alpha^{e/g} (p, \sigma) K^2 + O(K^4)
\end{equation}
with
\begin{equation}
\alpha^{e}(p, \sigma) =  2\,  \frac{3 + \left(\text{sinc}(\sigma) \right)^2 + 2 \left(\text{sinc}(\sigma/2) \right)^2
\left( 1 + \text{sinc}(\sigma) \right) + 4 \cos(2p) \left( \text{sinc}(\sigma)  + \left(\text{sinc}(\sigma/2) \right)^2 \right)
}{
3 + \left(\text{sinc}(\sigma) \right)^2 - 2 \left(\text{sinc}(\sigma/2) \right)^2
\left( 1 + \text{sinc}(\sigma) \right) + 4 \cos(2p) \left( \text{sinc}(\sigma)  - \left(\text{sinc}(\sigma/2) \right)^2 \right)
}
\label{eq:alphae}
\end{equation}
and 
\begin{equation}
\alpha^{g}(p, \sigma) =  2 \, \frac{1 + \left(\text{sinc}(\sigma)\right)^2}{1 - \left(\text{sinc}(\sigma)\right)^2}.
\label{eq:alphag}
\end{equation}
Note that $\alpha^g$ is actually independent of $p$.
%\footnote{Expression  (\ref{eq:alphag}) is actually only valid for $p \ne 0$. For DTQWs on the infinite line, $p$ takes all %values in $(-\pi, + \pi)$ and  (\ref{eq:alphag}) is val}.
Note also that the condition $\mid K \mid \ll \mid p \mid$ does not hinder asymptotic computations, at least on the infinite line. Indeed, as time increases, the density operator becomes more and more localized around $K = 0$, but it does not localize in $p$-space \footnote{A simple computation shows that the partly reduced density operator ${\hat \rho}^{pr}_j (p) = \sum_K {\hat \rho}_j(K, p)$ remains flat in $p$ at all times, as is the initial condition}. If one works on the infinite line, both $K$ and $p$ are continuous variables and the localization of the density operator around $K = 0$ implies that the size of the region in $p$-space where the condition $\mid K \mid \ll \mid p \mid$ does not apply actually shrinks to zero with time. For dynamics taking place on a finite circle (finite value of $M$), computations are a little more involved but can nevertheless be carried out. We feel a detailled analysis of the problem for finite values of $M$ does not bring any valuable insight on interesting physics or mathematics, and we thus restrict the analytical discussion of the asymptotic dunamics to DTQWs on the infinite line, where expressions (\ref{eq:alphae}) and (\ref{eq:alphag}) suffice.

A direct computation shows that the corrections to the eigenvectors are first order in $K$. By convention, we fix to unity the value of the first component of $w_1^{e/g}(K, p, \sigma)$ in the basis $(u_1, u_2, u_3, u_4)$. One thus gets for example 
\begin{equation}
w_1^{g}(K, p, \sigma) = \{\begin{matrix} 1, & \frac{2 i K \text{sinc}(\sigma)^2}{1-\text{sinc}(\sigma)^2}, & \frac{2 i K \text{sinc}(\sigma)}{1-\text{sinc}(\sigma)^2},& \frac{- 2 K \text{sinc}(\sigma) \tan(p)}{1-\text{sinc}(\sigma)^2} \end{matrix}\}.
\label{eq:w1g}
\end{equation}
The expression of $w_1^e$ is substantially more complicated and need not be reproduced here.
%Note finally that (\ref{eq:alphae}), (\ref{eq:alphag}) and (\ref{eq:w1g}) have been obtained from expansions in $K$
%valid for finite values of $p$. They are thus valid only if $\mid K \mid \ll \mid p \mid$. 

\section{Asymptotic expression of the density operator in Fourier space} 

Let us now use the above results to determine the time evolution of the average density operator in both cases under consideration. The first step is to express the initial condition, ${\hat \rho}_{j = 0} (K, p) = (u_1 - u_4)/2$ for all $(K, p)$, as a linear combination of the eigenvectors $w^{e/g}_r(K, p, \sigma)$. We thus write, for $a = 1, 2, 3, 4$
\begin{equation}
u_ a= \sum_{r = 1}^4 u^{e/g}_{ar} (K, p, \sigma) w^{e/g}_r(K, p, \sigma)
\end{equation}
and, conversely,
\begin{equation}
w^{e/g}_r(K, p, \sigma)= \sum_{a = 1}^4 w^{e/g}_{ra} (K, p, \sigma) u_a.
\end{equation}
By the above discussion of the eigenvalues and eigenvectors of ${\bar {\mathcal R}}^{e/g}_2$, one has notably $u^{e/g}_{11} (K, p, \sigma) = 1$,  $u^{e/g}_{1r} (K, p, \sigma) = O(K)$ for $r = 2, 3, 4$, $w^{e/g}_{11} (K, p, \sigma) = 1 + O(K)$.

One then writes`, for all $K$ and $p$:
%\footnote{From now on, the simplified notation $\hat \rho$ will be used instead of ${\bar {\hat \rho}}$}: 
\begin{equation}
{\hat \rho}_{j = 0} (K, p) = \frac{1}{2} 
\sum_{r = 1}^4 \left(u^{e/g}_{1r} (K, p, \sigma) - u^{e/g}_{4r} (K, p, \sigma) \right) w^{e/g}_r(K, p, \sigma).
\end{equation}
which leads to
\begin{equation}
{\hat \rho}_{j = J} (K, p) = \frac{1}{2} 
\sum_{r = 1}^4 \left( \lambda_r^{e/g} (K, p, \sigma) \right)^J \left(u^{e/g}_{1r} (K, p, \sigma) - u^{e/g}_{4r} (K, p, \sigma) \right) w^{e/g}_r(K, p, \sigma)
\end{equation}
or, expressing the eigenvectors $w^{e/g}_r(K, p, \sigma)$ in terms of the original basis vectors $(u_1, u_2, u_3, u_4)$:
\begin{equation}
{\hat \rho}_{j = J} (K, p) = \frac{1}{2} 
\sum_{r = 1}^4 \sum_{a = 1}^4  \left( \lambda_r^{e/g} (K, p, \sigma) \right)^J \left(u^{e/g}_{1r} (K, p, \sigma) - u^{e/g}_{4r} (K, p, \sigma) \right)  w^{e/g}_{ra} (K, p, \sigma) u_a
\label{eq:hatrhoNgene}
\end{equation}
Now, for all $r$, 
\begin{equation}
\lambda_r^{e/g} (K, p, \sigma)/ \lambda_1^{e/g} (K, p, \sigma) = \lambda_r^{e/g}(K =0, p, \sigma) \left( 1 + O(K^2)\right),
\end{equation}
since $\lambda_r^{e/g}(K =0, p, \sigma) = 1$. It follows that, for small enough $K$, the contributions to  (\ref{eq:hatrhoNgene}) proportional to $\left( \lambda_r^{e/g} (K, p, \sigma)\right)^J$ are much smaller than the 
contribution proportionnal to $(\lambda_1^{e/g} (K, p, \sigma))^J$ for all values of $p$ and $\sigma$ such that 
$\mid \lambda_r^{e/g}(K =0, p, \sigma)\mid < 1$. According to the above discussion, this is realized for all $r \ne 1$ and for all values of $p$ and $\sigma$, except in case 2 (random gravitational field) for $r = 4$, $p = \pm \pi$ or $p = 0$ and all values of $\sigma$. What happens at $p = \pm \pi$ has no incidence on the computation of the density operator in physical space. Indeed, for finite values of $M$, the maximum value $p_{\mbox{max}}$ of $\mid p \mid$ is $p_{\mbox{max}} = (2M/(2M + 1)) \pi < \pi$. Thus $\pm \pi$ is only reached in the limiting case of infinite $M$ {\sl i.e.} for quantum walks in the infinite line. However, $\pm p_{\mbox{max}} = \pm \pi$ then only appear as upper and lower bounds for integrals over $p$, and the values taken by ${\hat \rho} (J, K, p)$ at points $\pm \pi$ does not modify the values of the integrals. Moreover, all current computations are only valid for $\mid p \mid \ll \mid K \mid$ and are thus {sl a priori} invalid for $p = 0$. What happens around $p = 0$ has however no relevance to asymptotic computations on the infinite line because, as time increases, the density operator becomes more and more localized around $K = 0$ (see discussion below (\ref{eq:alphag})).

%The terms corresponding to $p = 0$ contribute to the expression of the density operator in physical space, but only for finite values of $M$ {\sl i.e.} on the circle. For quantum walks on the line, all summations over $K$ and $p$ are replaced by integrals and what happens at $p = 0$ does not modify the values of these integrals. For simplicity purposes, we thus choose now to restrict the discussion to quantum walks on the line and ignore the long-time contribution to the density operator at $p = 0$.

%In particular, the equivalent to equation (\ref{eq:hatrhoNe4}) is simply obtained by replacing $\alpha^e(p, sigma)

%\subsection{Random electric field}

%Let us first deal with case 1 i.e. transport in a random electric field. 

For large enough $J$ and small enough $K$, the double sum in (\ref{eq:hatrhoNgene}) thus simplifies into:
\begin{equation}
{\hat \rho}^{e/g}_{j = J}(K, p) = \frac{1}{2} 
\sum_{a = 1}^4  \left( \lambda_1^{e/g} (K, p, \sigma) \right)^J \left(u^{e/g}_{11} (K, p, \sigma) - u^{g/e}_{41} (K, p, \sigma) \right)  w^{e/g}_{1a} (K, p, \sigma) u_a
\label{eq:hatrhoNe1}
 \end{equation}
Now, $u_{11}^{e/g} (K, p, \sigma) = 1 + O(K)$, $u_{41}^{e/g} (K, p, \sigma) = O(K)$, $w_{11}^{e/g} (K, p, \sigma) = 1 + O(K)$ and $w_{1b}^{e/g} (K, p, \sigma) = O(K)$ for $b = 2, 3, 4$. As far as orders of magnitude are concerned, equation (\ref{eq:hatrhoNe1}) gives:
\begin{equation}
{\hat \rho}^{e/g}_{j = J} (K, p) = \frac{1}{2} 
\left( 1 - \alpha^{e/g} (p, \sigma) K^2 \right)^N \left(1 + O(K)\right) u_1 + \sum_{b = 2}^4 O(K) u_b.
\label{eq:hatrhoNe2}
 \end{equation}
At lowest order in $K$, $\left( 1 - \alpha^{e/g} (K, p, \sigma) K^2 \right)^J = 1 - \alpha^{e/g} (K, p, \sigma) J K^2$. We will now restrict the discussion to scales $K$ and times $J$ obeying $JK^2 \gg K$ i.e. $J K \gg 1$. Note that the maximum spatial spread of ${\bar \rho}$ at time $J$ is $L_{\text{max}}(J) = 2J$, so that the minimum value of $K$ for which ${\hat \rho}$ takes non negligible values at time $J$ is of order $K_{\text{min}}(N) = 1/J$. 
The condition $J K \gg 1$ thus restricts the discussion to length scales much smaller than $L_{\text{max}}(N)$. 
%Loosely speaking, the restriction encompasses the `center' of the density profiles, but not the edges. 
In particular, consider the time-dependent scale $K_J= K_*/\sqrt{J}$, where $K_*$ is an arbitrary time-independent wave-vector. The wave-vector $K_J$ obeys $JK_J^2 = K_*^2 \gg K_J$ for sufficiently large $J$. Thus, the possible diffusive behavior of the averaged transport is encompassed by the present discussion.

With the above assumption, equation (\ref{eq:hatrhoNe1}) implies the following approximate but very simple expression for the long time (large $J$) density operator in Fourier space:
\begin{equation}
{\hat \rho}^{e/g}_{j = J} (K, p) =  \frac{1}{2} 
\left( 1 - \alpha^{e/g} (p, \sigma) K^2 \right)^J u_1. 
\label{eq:hatrhoNe3}
 \end{equation}
In particular, for $K_J= K_*/\sqrt{J}$ (where $K_*$ is an arbitrary but $J$-independent wave number) and large enough $J$,
\begin{equation}
{\hat \rho}^{e/g}_{j = J} (K_J, p) =  \frac{1}{2} 
\left( 1 - \alpha^{e/g} (p, \sigma) \frac{K_*^2}{J} \right)^J u_1 
\sim  \frac{1}{2} \exp \left( - \alpha^{e/g} (p, \sigma) K_*^2\right) u_1
=  \frac{1}{2} \exp \left( - \alpha^{e/g} (p, \sigma) J K_J^2\right) u_1.
\label{eq:hatrhoNe4}
\end{equation}
This is the approximate expression for the asymptotic density operator presented in the main body of this article.

\bibliographystyle{unsrt}
\bibliography{bibli}

\begin{thebibliography}{10}

\bibitem{FeynHibbs65a}
R.P. Feynman and A.R. Hibbs.
\newblock Quantum mechanics and path integrals.
\newblock {\em International Series in Pure and Applied Physics. McGraw-Hill
  Book Company}, 1965.

\bibitem{ADZ93a}
Y.~Aharonov, L.~Davidovich, and N.~Zagury.
\newblock Quantum random walks.
\newblock {\em Phys. {R}ev. A}, 48:1687, 1993.

\bibitem{Meyer96a}
D.A. Meyer.
\newblock From quantum cellular automata to quantum lattice gases.
\newblock {\em {J}. {S}tat. {P}hys.}, 85, 1996.

\bibitem{Schmitz09a}
H.~Schmitz, R.~Matjeschk, Ch. Schneider, J.~Glueckert, M.~Enderlein, T.~Huber,
  and T.~Schaetz.
\newblock Quantum walk of a trapped ion in phase space.
\newblock {\em Phys. Rev. Lett.}, 103(090504):090504, August 2009.

\bibitem{Zahring10a}
F.~{Z\"a}hringer, G.~Kirchmair, R.~Gerritsma, E.~Solano, R.~Blatt, and C.F.
  Roos.
\newblock Realization of a quantum walk with one and two trapped ions.
\newblock {\em Phys. Rev. Lett.}, 104:100503, 2010.

\bibitem{Schreiber10a}
A.~Schreiber, K.N. Cassemiro, A.~G\'{a}bris V.~Poto\v{c}ek, P.J.Mosley,
  E.~Andersson, I.~Jex, and Ch. Silberhorn.
\newblock Photons walking the line.
\newblock {\em Phys. Rev. Lett.}, 104(050502):050502, 2010.

\bibitem{Karski09a}
Michal Karski, Leonid F\"{o}rster, Jai-Min Cho, Andreas Steffen, Wolfgang Alt,
  Dieter Meschede, and Artur Widera.
\newblock Quantum walk in position space with single optically trapped atoms.
\newblock {\em Science}, 325(5937):174--177, 2009.

\bibitem{Sansoni11a}
Sansoni L, Sciarrino F, Vallone G, Mataloni P, Crespi A, Ramponi R, and
  Osellame R.
\newblock Two-particle bosonic-fermionic quantum walk via 3d integrated
  photonics.
\newblock {\em Phys. Rev. Lett.}, 108(010502):010502, 2012.

\bibitem{Sanders03a}
B.C. Sanders, S.D. Bartlett, B.~Tregenna, and P.L. Knight.
\newblock Two-particle bosonic-fermionic quantum walk via 3d integrated
  photonics.
\newblock {\em Phys. {R}ev. A}, 67:042305, 2003.

\bibitem{Perets08a}
B.~Perets, Y.~Lahini, F.~Pozzi, M.~Sorel, R.~Morandotti, and Y.~Silberberg.
\newblock Realization of quantum walks with negligible decoherence in waveguide
  lattices.
\newblock {\em Phys. Rev. Lett.}, 100:170506, 2008.

\bibitem{var96a}
D.~Giulini, E.~Joos, C.~Kiefer, J.~Kupsch, I.-O. Stamatescu, and H.D. Zeh.
\newblock {\em Decoherence and the appearance of a Classical World in Quantum
  Theory}.
\newblock Springer-Verlag, Berlin, 1996.

\bibitem{Amb07a}
A.~Ambainis.
\newblock Quantum walk algorithm for element distinctness.
\newblock {\em SIAM Journal on Computing}, 37:210--239, 2007.

\bibitem{MNRS07a}
F.~Magniez, J.~Roland A.~Nayak, and M.~Santha.
\newblock Search via quantum walk.
\newblock {\em SIAM Journal on Computing - Proceedings of the thirty-ninth
  annual ACM symposium on Theory of computing}, New {Y}ork, 2007. ACM.

\bibitem{Aslangul05a}
C.~Aslangul.
\newblock Quantum dynamics of a particle with a spin-dependent velocity.
\newblock {\em Journal of Physics A: Mathematical and Theoretical}, 38:1--16,
  2005.

\bibitem{Bose03a}
S.~Bose.
\newblock Quantum communication through an unmodulated spin chain.
\newblock {\em Phys. Rev. Lett.}, 91:207901, 2003.

\bibitem{Burg06a}
D.~Burgarth.
\newblock Quantum state transfer with spin chains.
\newblock {\em University College London}, PhD thesis, 2006.

\bibitem{Bose07a}
S.~Bose.
\newblock Quantum communication through spin chain dynamics: an introductory
  overview.
\newblock {\em Contemp. {P}hys.}, 48(Issue 1):13 -- 30, January 2007.

\bibitem{Collini10a}
E.~Collini, C.Y. Wong, K.E. Wilk, P.M.G. Curmi, P.~Brumer, and G.D. Scholes.
\newblock {\em Nature}, page 644.

\bibitem{Engel07a}
G.S. Engel, T.R. Calhoun, R.L. Read, T.-K. Ahn, T.~Manal, Y.-C. Cheng, R.E.
  Blankenship, and G.~R. Fleming.
\newblock {\em Nature}, page 782.

\bibitem{DMD12a}
G.~Di Molfetta and F.~Debbasch.
\newblock Discrete-time quantum walks: Continuous limit and symmetries.
\newblock {\em J. {M}ath. {P}hys.}, 53:123302, 2012.

\bibitem{DMD13b}
G.~Di Molfetta, F.~Debbasch, and M.~Brachet.
\newblock Quantum walks as massless dirac fermions in curved space.
\newblock {\em Phys. Rev. A}, 88, 2013.

\bibitem{DMD14}
G.~Di Molfetta, F.~Debbasch, and M.~Brachet.
\newblock Quantum walks in artificial electric and gravitational fields.
\newblock {\em Phys. A}, 397, 2014.

\bibitem{Ken07a}
V.~Kendon.
\newblock Decoherence in quantum walks - a review.
\newblock {\em Math. Struct. in Comp. Sc.}, 17(6):1169--1220, 2007.

\bibitem{vieira14}
R.~Vieira, E.~P.~M. Amorim, and G.~Rigolin.
\newblock Entangling power of disordered quantum walks.
\newblock {\em Phys. Rev. A}, 89:042307, 2014.

\bibitem{brun2003}
Todd~A. Brun, Hilary~A. Carteret, and Andris Ambainis.
\newblock Quantum to classical transition for random walks.
\newblock {\em Phys. Rev. Lett.}, 91:130602, Sep 2003.

\bibitem{Werner11}
A.~Ahlbrecht, H.~Vogts, A.~H. Werner, and R.~F. Werner.
\newblock Asymptotic evolution of quantum walks with random coin.
\newblock {\em Journal of Mathematical Physics}, 52(4), 2011.

\bibitem{Cedzich12}
Andre Ahlbrecht, Christopher Cedzich, Robert Matjeschk, VolkherB. Scholz,
  AlbertH. Werner, and ReinhardF. Werner.
\newblock Asymptotic behavior of quantum walks with spatio-temporal coin
  fluctuations.
\newblock {\em Quantum Information Processing}, 11(5):1219--1249, 2012.

\bibitem{Joye11}
Alain Joye.
\newblock Random time-dependent quantum walks.
\newblock {\em Communications in Mathematical Physics}, 307(1):65--100, 2011.

\bibitem{kollar14}
B.~Kollar and M.~Koniorczyk.
\newblock Entropy rate of message sources driven by quantum walks.
\newblock {\em Phys. Rev. A}, 89:022338, 2014.

\bibitem{liu10}
Chaobin Liu and Nelson Petulante.
\newblock On the von neumann entropy of certain quantum walks subject to
  decoherence.
\newblock {\em Mathematical Structures in Computer Science}, 20:1099--1115, 12
  2010.

\bibitem{chandra10a}
C.M. Chandrasekhar, S.~Banerjee, and R.~Srikanth.
\newblock Relationship between quantum walks and relativistic quantum
  mechanics.
\newblock {\em Phys. {R}ev. A}, 81:062340, 2010.

\bibitem{abal06}
G.~Abal, R.~Siri, A.~Romanelli, and R.~Donangelo.
\newblock Quantum walk on the line: Entanglement and nonlocal initial
  conditions.
\newblock {\em Phys. Rev. A}, 73:042302, Apr 2006.

\bibitem{chisaki2011}
Kota Chisaki, Norio Konno, Etsuo Segawa, and Yutaka Shikano.
\newblock Crossovers induced by discrete-time quantum walks.
\newblock {\em Quantum Information \& Computation}, 11(9-10):741--760, 2011.

\bibitem{perez10}
C.~Navarrete-Benlloch, A.~Perez, and Eugenio Roldan.
\newblock Nonlinear optical galton board.
\newblock {\em Phys. Rev. A}, 75:062333, 2010.

\bibitem{obuse2011topological}
Hideaki Obuse and Norio Kawakami.
\newblock Topological phases and delocalization of quantum walks in random
  environments.
\newblock {\em Physical Review B}, 84(19):195139, 2011.

\bibitem{shikano2014discrete}
Yutaka Shikano, Tatsuaki Wada, and Junsei Horikawa.
\newblock Discrete-time quantum walk with feed-forward quantum coin.
\newblock {\em Scientific reports}, 4, 2014.

\bibitem{prokof2006decoherence}
NV~Prokof?ev and PCE Stamp.
\newblock Decoherence and quantum walks: Anomalous diffusion and ballistic
  tails.
\newblock {\em Physical Review A}, 74(2):020102, 2006.

\bibitem{schreiber2011decoherence}
A~Schreiber, KN~Cassemiro, V~Poto{\v{c}}ek, A~G{\'a}bris, I~Jex, and
  Ch~Silberhorn.
\newblock Decoherence and disorder in quantum walks: From ballistic spread to
  localization.
\newblock {\em Physical review letters}, 106(18):180403, 2011.

\bibitem{inui2004localization}
Norio Inui, Yoshinao Konishi, and Norio Konno.
\newblock Localization of two-dimensional quantum walks.
\newblock {\em Physical Review A}, 69(5):052323, 2004.

\bibitem{D04a}
F.~Debbasch.
\newblock What is a mean gravitational field?
\newblock {\em Eur.~Phys.~J.~B}, 37(2):257--270, 2004.

\bibitem{CaldLegg87}
A.~J. Leggett, S.~Chakravarty, A.~T. Dorsey, Matthew P.~A. Fisher, Anupam Garg,
  and W.~Zwerger.
\newblock Dynamics of the dissipative two-state system.
\newblock {\em Rev. Mod. Phys.}, 59:1--85, Jan 1987.

\bibitem{Cald83}
A.O. Caldeira and A.J. Leggett.
\newblock Path integral approach to quantum brownian motion.
\newblock {\em Physica A: Statistical Mechanics and its Applications},
  121(3):587 -- 616, 1983.

\bibitem{DMR97a}
F.~Debbasch, K.~Mallick, and J.P. Rivet.
\newblock Relativistic {Ornstein}-{Uhlenbeck} process.
\newblock {\em J. Stat. Phys.}, 88:945, 1997.

\bibitem{CD07g}
C.~Chevalier and F.~Debbasch.
\newblock Relativistic diffusions: a unifying approach.
\newblock {\em J.~{M}ath.~{P}hys.}, 49:043303, 2008.

\bibitem{DEF12a}
F.~Debbasch, D.~Espaze, and V.~Foulonneau.
\newblock Can diffusions propagate?
\newblock {\em J.Stat.Phys}, 149:37--49, 2012.

\bibitem{DMD13a}
G.~Di Molfetta and F.~Debbasch.
\newblock Discrete-time quantum walks: Continuous limit in 1 + 1 and 1 + 2
  dimension.
\newblock {\em J.Comp.Th.Nanosc.}, 10,7:1621--1625, 2012.

\bibitem{DA14}
P.~Arnault and F.~Debbasch.
\newblock Landau levels for discrete-time quantum walks in artificial magnetic
  fields.
\newblock {\em arXiv preprint}, 1412.4337, 2014.

\bibitem{W84a}
R.M. Wald.
\newblock {\em General Relativity}.
\newblock The University of Chicago Press, Chicago, 1984.

\end{thebibliography}

\end{document}